\documentclass[sigconf]{acmart}

\usepackage[ruled,linesnumbered]{algorithm2e}
\usepackage{tikz}
\usetikzlibrary{positioning, shapes.geometric, arrows}
\tikzstyle{process} = [rectangle, rounded corners, minimum width=3cm, minimum height=1cm,text centered, draw=black, fill=gray!10]
\tikzstyle{database} = [cylinder, shape border rotate=90, minimum height=2cm, minimum width=1cm, text centered, draw=black, fill=gray!20]
\tikzstyle{decision} = [diamond, minimum width=3cm, minimum height=1cm, text centered, draw=black, fill=gray!20]
\tikzstyle{arrow} = [thick,->,>=stealth]

\AtBeginDocument{%
  }

\copyrightyear{2024}
\acmYear{2024}
\acmISBN{978-1-4503-XXXX-X/18/06}

\begin{document}

\title{FRAG: Toward Federated Vector Database Management for Collaborative and Secure Retrieval-Augmented Generation}

\author{Dongfang Zhao}
\affiliation{%
  \institution{University of Washington}
  \country{dzhao@cs.washington.edu}
}

\newcommand{\don}[1]{\textcolor{magenta}{Dongfang: #1}}

\begin{abstract}
This paper introduces \textit{Federated Retrieval-Augmented Generation (FRAG)}, a novel database management paradigm tailored for the growing needs of retrieval-augmented generation (RAG) systems, which are increasingly powered by large-language models (LLMs). FRAG enables mutually-distrusted parties to collaboratively perform Approximate $k$-Nearest Neighbor (ANN) searches on encrypted query vectors and encrypted data stored in distributed vector databases, all while ensuring that no party can gain any knowledge about the queries or data of others.
Achieving this paradigm presents two key challenges: 
(i) ensuring strong security guarantees, such as Indistinguishability under Chosen-Plaintext Attack (IND-CPA), under practical assumptions (e.g., we avoid overly optimistic assumptions like non-collusion among parties); and 
(ii) maintaining performance overheads comparable to traditional, non-federated RAG systems. 
To address these challenges, FRAG employs a single-key homomorphic encryption protocol that simplifies key management across mutually-distrusted parties. Additionally, FRAG introduces a \textit{multiplicative caching} technique to efficiently encrypt floating-point numbers, significantly improving computational performance in large-scale federated environments.
We provide a rigorous security proof using standard cryptographic reductions and demonstrate the practical scalability and efficiency of FRAG through extensive experiments on both benchmark and real-world datasets.

\end{abstract}

\received{20 February 2007}
\received[revised]{12 March 2009}
\received[accepted]{5 June 2009}

\settopmatter{printfolios=true}
\maketitle

\section{Introduction}

\subsection{Background}
In recent years, large-language models (LLMs) such as ChatGPT~\cite{openai2023gpt4}, LLaMA~\cite{touvron2023llama}, BERT~\cite{devlin2018bert}, and Transformer~\cite{raffel2020t5} have fundamentally transformed the landscape of natural language processing (NLP) and artificial intelligence (AI). These models, trained on vast amounts of textual data, have demonstrated unprecedented capabilities in generating human-like text, understanding complex queries, and performing tasks that require deep contextual understanding~\cite{brown2020language, devlin2018bert}. Applications leveraging LLMs span across numerous industries, from automated customer service and real-time translation to content creation and medical diagnostics. The flexibility and power of LLMs have made them central to many modern AI systems.

A crucial enhancement in the effectiveness of LLMs has been the integration of \textit{Retrieval-Augmented Generation (RAG)}. Unlike purely generative models, RAG systems enhance their output by retrieving relevant external knowledge from vast databases or knowledge bases to augment their responses~\cite{lewis2020retrieval}. This mechanism allows LLMs to generate more informed, accurate, and contextually rich outputs by incorporating real-time or pre-stored information. For example, when answering factual queries, the model can retrieve up-to-date knowledge rather than relying solely on its pre-trained information. As a result, RAG has become a vital component in applications where the accuracy of the generated content is paramount, such as in recommendation systems, conversational agents, and information retrieval.

At the heart of RAG systems are vector embeddings, which represent pieces of data, such as words, sentences, or entire documents, as high-dimensional vectors. These embeddings capture semantic similarities between different entities, enabling efficient retrieval of related data. Vector databases like Faiss~\cite{johnson2019billion}, Annoy~\cite{bernardini2017annoy}, and HNSW~\cite{malkov2018efficient} have been widely adopted for storing and querying these embeddings, offering optimized solutions for Approximate $k$-Nearest Neighbor (ANN) searches on large-scale datasets~\cite{johnson2019billion, bernardini2017annoy, malkov2018efficient}. These systems rely on specialized indexing and search algorithms to quickly identify the most relevant embeddings from millions or billions of records, making them indispensable in modern AI pipelines.

\subsection{Motivation}

Despite the centralized nature of existing RAG systems, organizations across various industries recognize the potential benefits of collaboration. By pooling their data resources, organizations can improve the performance of machine learning models, gain access to a broader knowledge base, and make better-informed decisions. For instance, in the healthcare sector, multiple hospitals could collaborate to build a more comprehensive model for diagnosing diseases by leveraging each hospital's patient data. In finance, organizations might wish to jointly assess risk using data from different institutions. By sharing insights, these entities can significantly enhance the accuracy and robustness of their models, ultimately improving their services and outcomes~\cite{bonawitz2017practical}.

However, in industries like healthcare, finance, and legal services, datasets often contain sensitive or proprietary information that cannot be shared freely due to privacy regulations (e.g., GDPR and HIPAA) and competitive concerns. Despite the potential benefits of collaborative AI, privacy concerns and regulatory constraints often prevent organizations from directly sharing their raw data. Therefore, the need for privacy-preserving techniques that allow secure collaboration without compromising sensitive data becomes paramount.

This challenge is particularly acute in scenarios involving mutually-distrusted parties. For example, in a federated environment where multiple organizations wish to collaborate on training or querying a model, none of the participants are willing to expose their raw data to others. At the same time, these organizations need to perform collaborative retrieval tasks, such as ANN searches across distributed datasets. The key challenge is enabling such tasks without compromising the privacy of any party's data or queries. Traditional privacy-preserving solutions, such as Secure Multi-Party Computation (MPC)~\cite{mohassel2017secureml} and Homomorphic Encryption (HE)~\cite{gentry2009fully}, while theoretically robust, often suffer from high computational overheads and complex key management. These issues become particularly pronounced when applied to real-time RAG systems that require fast, scalable solutions to handle the massive volumes of data typically involved.

\subsection{Proposed Work}

To address the challenges of secure collaboration in Retrieval-Augmented Generation (RAG) systems, we propose \textit{Federated Retrieval-Augmented Generation (FRAG)}, a novel framework that enables distributed, privacy-preserving RAG in federated environments. The FRAG framework allows mutually-distrusted parties to perform collaborative Approximate $k$-Nearest Neighbor (ANN) searches on encrypted query vectors and encrypted data stored across distributed vector databases, ensuring that no party can infer the data or queries of others. Importantly, FRAG provides strong security guarantees while maintaining performance efficiency close to that of centralized, non-federated RAG systems.

The key innovation in FRAG is its use of the \textit{Single-Key Homomorphic Encryption (SK-MHE)} protocol. Unlike traditional multi-key homomorphic encryption schemes, SK-MHE simplifies key management by allowing all parties to use a single shared encryption key. This reduces the complexity of key exchanges while ensuring security against a wide range of attacks, including Indistinguishability under Chosen-Plaintext Attack (IND-CPA). By leveraging SK-MHE, FRAG ensures that encrypted vector operations, such as scalar products and distance computations for ANN searches, can be performed securely without compromising data privacy.

In addition, FRAG introduces a \textit{Multiplicative Caching (MC)} protocol to optimize the computational efficiency of homomorphic operations. Since vector embeddings often involve floating-point arithmetic, MC precomputes and caches intermediate encrypted values, which can then be reused during subsequent computations. This significantly reduces the computational burden typically associated with homomorphic operations, making FRAG practical for large-scale, real-time ANN searches.

FRAG is specifically designed to address the performance bottlenecks and security concerns in federated RAG environments. Through the integration of SK-MHE and MC, FRAG provides an efficient and scalable solution for performing encrypted ANN searches across distributed databases, making it applicable to a wide range of real-world use cases, such as collaborative healthcare research, federated financial analysis, and secure multi-institutional machine learning.

\subsection{Contributions}

The key contributions of this paper are as follows:
\begin{itemize}
    \item We propose the \textit{Federated Retrieval-Augmented Generation (FRAG)} framework, which allows mutually-distrusted parties to perform collaborative ANN searches on encrypted data while preserving privacy.
    \item We introduce the \textit{Single-Key Homomorphic Encryption (SK-MHE)} protocol, which simplifies key management in federated environments while maintaining strong security guarantees, including IND-CPA security.
    \item We develop a \textit{Multiplicative Caching (MC)} protocol that optimizes the performance of homomorphic operations by caching intermediate encrypted values, reducing computational overhead for large-scale vector operations.
    \item We provide a rigorous security analysis and demonstrate the scalability and efficiency of FRAG through extensive experiments on synthetic and real-world datasets, showing that it achieves performance comparable to centralized RAG systems while maintaining privacy guarantees.
\end{itemize}

\section{Related Work}

\subsection{Vector Databases}

Vector databases have gained prominence with the rise of applications such as natural language processing, recommendation systems, and information retrieval, where large-scale vector-based similarity searches are essential. Various systems have been developed to cater to these needs. For example, systems like EuclidesDB, Pinecone, and Vearch focus on efficient vector search but often lack flexibility in handling non-vector queries or mixed workloads~\cite{milvus2021, euclidesDB, pinecone2019}. These systems primarily employ approximate nearest neighbor (ANN) search techniques, with graph-based methods such as Hierarchical Navigable Small Worlds (HNSW) being particularly popular for scaling large datasets~\cite{malkov2018, johnson2019}.

Milvus, a purpose-built vector database management system, stands out by supporting multiple index types and hybrid queries, where structured attributes and vector-based similarity searches are combined~\cite{milvus2021}. This allows for high flexibility, catering to both traditional query optimization techniques and emerging data types like unstructured text and images. Other systems, such as Weaviate and Qdrant, integrate more advanced features like predicated vector search and cost-based optimization, providing a more comprehensive framework for federated and distributed environments~\cite{qdrant2021, weaviate2021}.

However, many existing solutions lack efficient mechanisms to address privacy concerns in federated learning or distributed settings. This challenge is especially relevant for real-time systems where cryptographic protocols could impose significant computational overhead~\cite{secureml2016, privacyFL2019}. While previous works such as privacy-preserving machine learning frameworks like SecureML provide foundational approaches, our protocol aims to optimize vector data management by introducing secure and efficient cryptographic protocols for distributed environments, improving both performance and privacy in federated settings~\cite{secureml2016}.

\subsection{Federated Learning and Retrieval}

Federated learning~\cite{fedlearn} has emerged as a popular paradigm for enabling decentralized machine learning models, where data remains local to each party while a global model is collaboratively trained. Similarly, federated search~\cite{fedsearch} focuses on enabling multiple parties to perform retrieval tasks without sharing raw data. While federated learning and retrieval have gained significant attention, most existing methods are designed for traditional machine learning or document retrieval systems and do not address the unique requirements of vector databases, especially in RAG systems.

Federated vector databases are an underexplored area. Early work has primarily focused on the basic functionality of federated data retrieval without considering privacy and security concerns in depth. FRAG fills this gap by enabling secure and efficient vector search in a federated setting, addressing both privacy and performance challenges.

\subsection{Secure Aggregation and Poisoning Attacks}

Secure aggregation algorithms often assume a centralized server for effective collaboration. In a study by Yin et al.~\cite{dyin_icml18}, robust distributed gradient descent algorithms utilizing median and trimmed mean operations were analyzed. Additionally, Lyu et al.~\cite{lli_aaai19} introduced robust stochastic sub-gradient methods for distributed learning from heterogeneous datasets, addressing the challenge of adversarial participants. These algorithms typically operate under the \textit{semi-honest} model, wherein participants adhere to the protocol but may passively analyze data~\cite{zhang_atc20}. This model is widely used in production systems.

Beyond the semi-honest model, a more aggressive assumption involves \textit{malicious} participants, who deviate from the protocol and manipulate data. For instance, Alfeld et al.~\cite{alfeld2016} discussed \textit{data poisoning} attacks, where compromised samples are injected into the dataset. Similarly, \textit{model poisoning} attacks involve uploading tampered models~\cite{li2020review}. Well-known examples include backdoor attacks, as studied in~\cite{yang2019, bhagoji2019analyzing, BagdasaryanVHES20, Xie2020DBA}. In response, the security community has proposed diverse solutions to defend against these poisoning attacks, including detection and mitigation methods~\cite{mfang_security20, ddata_icml21, yyang_icml21, sun2021, Shejwalkar2021}.

\subsection{Homomorphic Encryption and Secret Sharing}

Homomorphic encryption is a key technique in secure computations, enabling operations on encrypted data without decryption. Gentry’s work~\cite{cgentry_stoc09} introduced fully homomorphic encryption (FHE), which supports both additive and multiplicative homomorphisms. However, FHE schemes like BFV~\cite{bfv} and CKKS~\cite{ckks} impose significant computational overhead, making them less practical for many real-time applications. In contrast, more efficient schemes like Paillier encryption~\cite{ppail_eurocrypt99} have been widely applied in privacy-preserving machine learning~\cite{symmetria_vldb20, shardy_arxiv17, zhang_atc20}.

An alternative to cryptographic approaches is threshold secret sharing (TSS)~\cite{ashamir_cacm79, trabin_stoc89}, which splits data into $n$ shares, ensuring that only a subset of $t$ or more shares can reconstruct the original message. TSS plays an important role in secure multi-party computation (MPC) and has been incorporated into various frameworks such as DeepSecure~\cite{deepsecure}, SecureML~\cite{secureml}, and ABY~\cite{aby}, enabling privacy-preserving collaborative computations.

\subsection{Multi-Party Computation}

Multi-Party Computation (MPC) allows multiple parties to jointly compute a function over their inputs while keeping those inputs private. Classic works such as Yao’s Garbled Circuits~\cite{yao}, and more recently, homomorphic encryption schemes~\cite{gentry} have laid the foundation for privacy-preserving computations. MPC protocols have been widely adopted in privacy-sensitive applications, but they often face limitations in terms of scalability and performance, particularly when applied to complex tasks like ANN search in high-dimensional vector spaces.

Recent work in homomorphic encryption (HE) and differential privacy~\cite{he_dp} has aimed to mitigate some of these performance bottlenecks. However, these approaches typically involve high computational costs, especially for floating-point operations. Our contribution in FRAG leverages a single-key homomorphic encryption protocol, which simplifies key management, and introduces the novel \textit{multiplicative caching} technique to reduce the overhead associated with floating-point computations.

\subsection{Provable Security}

When deploying an encryption scheme, proving its security is crucial. Provable security involves specifying the security goal, threat model, and assumptions. One well-known security goal is to achieve resistance against \textit{Chosen-Plaintext Attacks} (CPA)~\cite{cgentry_stoc09}, where an adversary can query plaintext-ciphertext pairs and still fail to decrypt new ciphertexts. Various encryption schemes, including Paillier~\cite{ppail_eurocrypt99} and FHE~\cite{cgentry_stoc09}, are designed to be CPA-secure. In cryptographic proofs, \textit{negligible functions} quantify the probability of adversarial success, ensuring that the attack probability decreases exponentially with the key size~\cite{ppail_eurocrypt99, cgentry_stoc09}.

The field of vector database management and retrieval has seen rapid advancements, particularly with the increasing demand for high-performance, large-scale systems capable of handling massive datasets. This section reviews key areas related to our work: vector databases, federated learning, multi-party computation (MPC), and homomorphic encryption.

\section{Problem Statement}

\subsection{System Model}

We consider a federated environment comprising $n$ mutually-distrusted parties $P_1, P_2, \dots, P_n$, each maintaining its own encrypted vector database. The data in each party’s database is composed of high-dimensional vectors, $D_i = \{ \mathbf{v}_{i1}, \mathbf{v}_{i2}, \dots, \mathbf{v}_{im} \}$, where $i$ represents the party index, and $m$ is the number of vectors in the database. These vectors may represent a variety of data types, such as text embeddings, image features, or user profiles, depending on the use case.

Each party also holds a set of encrypted query vectors, $\mathbf{q}_i$, and wishes to perform an ANN search across the encrypted databases of all parties in the system. However, no party wants to disclose its query vectors or database contents to others. Our goal is to allow this federated ANN search to be executed in a privacy-preserving manner, while ensuring high performance comparable to plaintext systems.

\subsection{Threat Model}

The primary threat in this federated setup is that parties may try to infer information about other parties’ data during the query and retrieval process. We assume a semi-honest (honest-but-curious) model, where parties follow the prescribed protocol but may attempt to gather additional information from the data they receive. This model is common in federated learning and secure data-sharing environments, where parties agree to collaborate under the assumption that others may not be fully trusted.

Additionally, we do not assume non-collusion among parties. This means two or more parties could collude to extract more information than they could independently. Therefore, our system’s security guarantees must hold even in the presence of collusion among a subset of parties.

\subsection{Research Problem}

Given the above system model, we formalize the core research problem as follows:

\begin{itemize}
    \item \textbf{Federated ANN Search}: Given a query vector $\mathbf{q}_i$ from party $P_i$, the system must perform an Approximate $k$-Nearest Neighbor (ANN) search across the encrypted databases of all parties $P_1, P_2, \dots, P_n$, and return the $k$ closest matches without revealing any party’s data.
    
    \item \textbf{Security Requirement}: Ensure that no party $P_i$ gains access to any information about other parties’ vectors or queries during the entire process, except for the final ANN result. Specifically, the system must achieve security guarantees, such as Indistinguishability under Chosen-Plaintext Attack (IND-CPA).
    
    \item \textbf{Performance Requirement}: Minimize the computational and communication overhead associated with encryption, query processing, and result aggregation, so that the performance of FRAG is comparable to that of a centralized, plaintext-based ANN system.
\end{itemize}

\subsection{Technical Challenges}

Achieving these goals presents several technical challenges:

\begin{itemize}
    \item \textbf{Secure Homomorphic Operations}: The core task of performing an ANN search on encrypted data requires the execution of homomorphic operations, which are typically computationally expensive. The challenge lies in designing a system that can perform these operations efficiently while maintaining strong security guarantees.
    
    \item \textbf{Key Management}: In a federated environment with mutually-distrusted parties, managing cryptographic keys becomes a complex problem. The system must ensure that each party can securely encrypt and decrypt its own data with a simplified and scalable key management scheme, without compromising the overall security.
    
    \item \textbf{Performance Optimization}: Homomorphic encryption often incurs significant overhead in both computation and communication. Optimizing the encryption, query processing, and result aggregation steps to reduce this overhead is critical to making FRAG practical for real-world use in large-scale federated environments.
\end{itemize}

\section{FRAG Overview}

\subsection{System Architecture}

\begin{figure}[t]
    \centering
    \includegraphics[width=100mm]{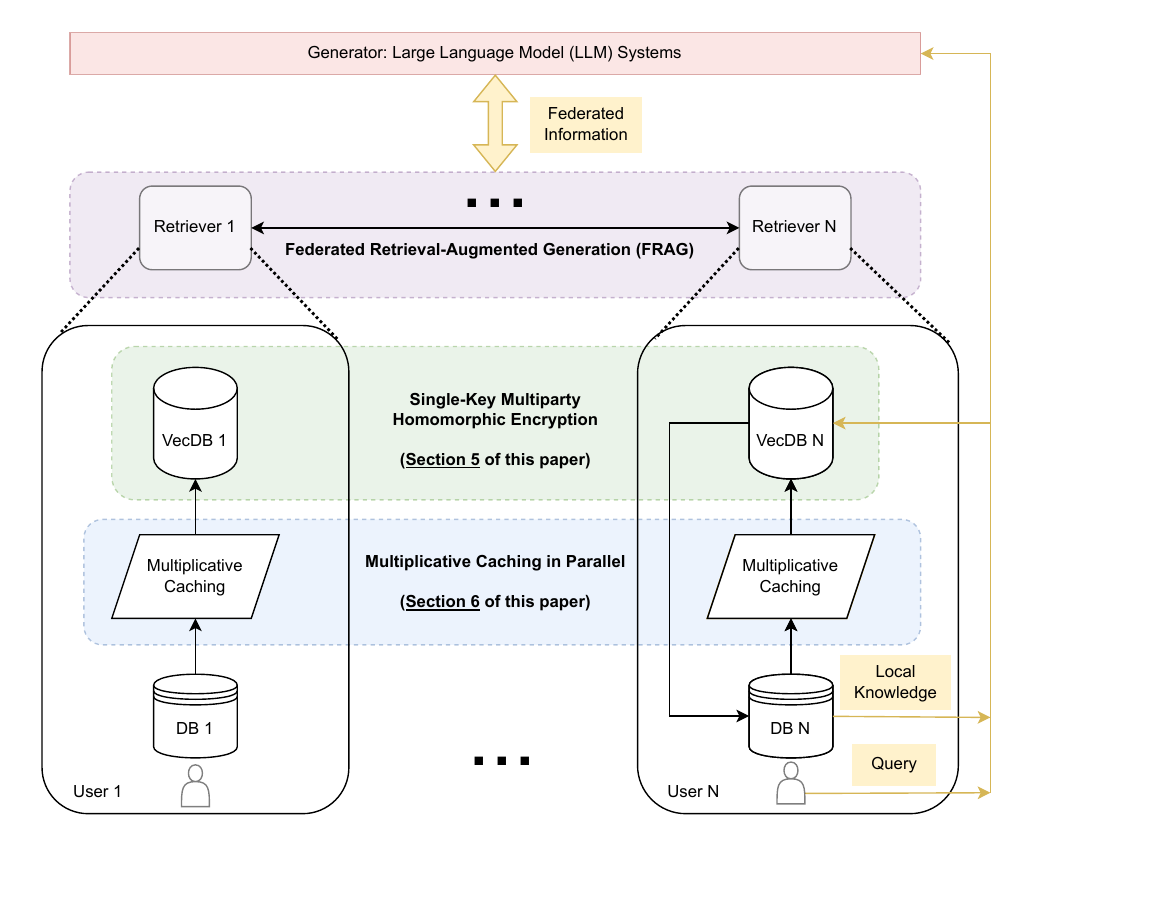}
    \caption{FRAG Architecture}
    \label{fig:arch}
\end{figure}

The overall architecture of the Federated Retrieval-Augmented Generation (FRAG) system is depicted in Figure~\ref{fig:arch}. FRAG consists of multiple distributed nodes, each maintaining a local vector database (VecDB), enabling secure and collaborative Approximate $k$-Nearest Neighbor (ANN) searches. These searches leverage advanced cryptographic techniques like \textit{Single-Key Multiparty Homomorphic Encryption (SK-MHE)} and \textit{Multiplicative Caching (MC)} to protect the privacy of data throughout the process.

Each FRAG node is structured around two primary components:

\paragraph{Local Vector Database (VecDB)} 
Each participating node stores local, high-dimensional embeddings in a secure vector database (VecDB). These embeddings represent data used for retrieval tasks like ANN searches. VecDB ensures that all stored data vectors are encrypted using the \textit{SK-MHE protocol}, preserving privacy during federated computations.

\paragraph{Single-Key Multiparty Homomorphic Encryption (SK-MHE)}
The \textit{SK-MHE protocol} provides secure, homomorphic operations on encrypted vectors across multiple federated nodes. By sharing a single encryption key across participating nodes, SK-MHE facilitates collaborative, encrypted computations—such as dot-product calculations and distance evaluations—without decrypting the underlying data. This ensures full privacy during query and retrieval operations.

\paragraph{Multiplicative Caching (MC)}
The \textit{Multiplicative Caching} protocol is designed to reduce the computational overhead associated with homomorphic operations. Precomputed scalar values, essential for vector operations, are cached in encrypted form. This allows each node to quickly perform encrypted scalar and vector multiplications during ANN searches, improving efficiency, especially in large-scale federated environments.

\paragraph{Query Distribution and Aggregation}
In FRAG, encrypted queries are distributed across multiple nodes, each performing secure retrieval operations using the SK-MHE protocol. The results are encrypted and returned to a central aggregator, which combines the encrypted results and sends them back to the querying node. The querying node decrypts the results using the shared key, thus maintaining privacy throughout the entire process.

\subsection{Query Workflow}

The workflow in FRAG ensures secure, distributed data retrieval while leveraging advanced homomorphic encryption techniques. The steps are detailed as follows:

\paragraph{Step 1: Query Submission}
A user submits an encrypted query to the FRAG system. The query vector is encrypted using the shared key provided by the \textit{SK-MHE protocol}, ensuring that no sensitive data is exposed during transmission or computation.

\paragraph{Step 2: Query Distribution}
The encrypted query is distributed to multiple nodes, each equipped with a local VecDB. Each node retrieves the most relevant encrypted vectors from its database, prepared for further encrypted computations.

\paragraph{Step 3: Secure Retrieval with SK-MHE}
Each node performs an encrypted ANN search on the local vectors using the \textit{SK-MHE protocol}. Homomorphic encryption ensures that both the query vector and the data vectors remain confidential throughout the process.

\paragraph{Step 4: Performance Optimization via Multiplicative Caching}
During the ANN search, the \textit{Multiplicative Caching} protocol accelerates the homomorphic computations. Cached encrypted scalar products reduce the need for repeated encryptions, enhancing the speed of operations across the federated nodes.

\paragraph{Step 5: Encrypted Result Transmission}
After performing the ANN search, each node encrypts its local results and transmits them to the central aggregator. The results remain encrypted during the transmission, protecting the data from exposure.

\paragraph{Step 6: Aggregation and Decryption}
The central aggregator collects the encrypted results from all nodes and combines them into a single encrypted response. This response is returned to the querying user, who decrypts the result using the shared key.

\paragraph{Step 7: Integration with LLM Systems}
FRAG supports integration with \textit{Large Language Model (LLM) systems}. Once decrypted, the results can be fed into LLMs to generate context-aware, augmented responses, all while preserving the privacy of user data.

\section{Single-Key Multiparty Homomorphic Encryption (SK-MHE)}
\label{sec:skmhe}

\subsection{Protocol Description}

The Single-Key Homomorphic Encryption (SK-MHE) protocol is designed to allow mutually-distrusted parties to securely compute over encrypted data while preventing unauthorized access to the underlying information. A crucial element of this protocol is the ciphertext splitting mechanism, which ensures that no single party can decrypt the data without collaboration. The process begins with the encryption of a data vector $\mathbf{v}$ using a shared public key $pk$. The encryption operation produces a ciphertext $\mathbf{c}$ that is securely linked to the original data. 

Once the data is encrypted, the ciphertext is split into multiple shares, denoted as $\{ s_1, s_2, \dots, s_n \}$, where $n$ is the number of participating parties. Each share represents a fragment of the original ciphertext, and none of the parties can independently reconstruct the full ciphertext from a single share. The splitting process ensures that the encrypted data is distributed securely, making it impossible for any one party to recover the plaintext without all the shares.

After the shares are generated, they are distributed to the distinct parties involved in the protocol. Each party only holds one share, and as a result, no single entity has enough information to decrypt the data independently. This distributed approach protects the confidentiality of the data, even in the event of partial compromise.

When a query is issued, the querying party provides an encrypted query vector $\mathbf{q}$. Each party performs homomorphic computations on its own share of the ciphertext and the encrypted query. This step involves secure operations on the data that remain entirely within the encrypted domain, ensuring that the original data and the query are never exposed. The results of these computations are intermediate values corresponding to each share. These intermediate results are kept secure through encryption.

Once all parties have completed their homomorphic computations, the results are aggregated. The aggregation step combines the intermediate results into a single encrypted output, while still preserving the security guarantees provided by the encryption. At no point during the aggregation process is any plaintext revealed, ensuring that the entire computation remains confidential.

Finally, the querying party, who holds the private key $sk$, decrypts the aggregated result. This decryption step reveals the final result of the computation. Since only the querying party has access to the private key, the other parties cannot access the decrypted result. Throughout the entire process, from encryption to decryption, the security of the data is preserved, ensuring that sensitive information is never exposed to unauthorized entities.

We sketch the SK-MHE protocol in Alg.~\ref{alg:sk-mhe},
which leverages the combined strength of homomorphic encryption and ciphertext splitting to provide both security and efficiency in federated environments. By ensuring that no single party can access the full ciphertext or plaintext, it maintains the privacy of sensitive data while allowing for collaborative computations.

\begin{algorithm}[!t]
\SetAlgoLined
\caption{SK-MHE Protocol}
\label{alg:sk-mhe}
\KwIn{$\mathbf{v} = (v_1, v_2, \dots, v_m)$, $\mathbf{q} = (q_1, q_2, \dots, q_m)$, $n$, $pk$, $sk$}
\KwOut{$\mathbf{v}_{\text{result}}$}

$\mathbf{c} = (c_1, c_2, \dots, c_m) \gets (Enc(v_1, pk), Enc(v_2, pk), \dots, Enc(v_m, pk))$

\For{$i = 1 \dots m$}{
    $\{ s_1^i, s_2^i, \dots, s_n^i \} = Split(c_i, n)$
}

$\mathbf{q}_E = (Enc(q_1, pk), Enc(q_2, pk), \dots, Enc(q_m, pk))$

\For{$i = 1 \dots m$}{
    \For{$j = 1 \dots n$}{
        $r_j^i \gets HomomorphicOp(s_j^i, Enc(q_i, pk))$
    }
}

\For{$j = 1 \dots n$}{
    $R_j = \sum_{i=1}^{m} r_j^i$
}

$\text{result} = \sum_{j=1}^{n} R_j$

$\text{result}_{dec} = Dec(\text{result}, sk)$

$\mathbf{v}_{\text{result}} = (v_{\text{res1}}, v_{\text{res2}}, \dots, v_{\text{resm}}) \gets \mathbf{q}_E^{-1} \times \text{result}_{dec}$

\For{$i = 1 \dots m$}{
    $\text{Verify}(v_{\text{res}_i})$ \quad \text{if valid, continue; otherwise, recompute}
}

\Return $\mathbf{v}_{\text{result}}$
\end{algorithm}

\subsection{Correctness Analysis}

The correctness of the SK-MHE protocol relies on the properties of homomorphic encryption and the secure aggregation of encrypted shares. Given a data vector $\mathbf{v} = (v_1, v_2, \dots, v_m)$ and a query vector $\mathbf{q} = (q_1, q_2, \dots, q_m)$, the SK-MHE protocol guarantees that the final decrypted result corresponds to the correct computation on the plaintext.

Each element of $\mathbf{v}$ is first encrypted using the public key $pk$, yielding a ciphertext vector $\mathbf{c} = (Enc(v_1, pk), Enc(v_2, pk), \dots, Enc(v_m, pk))$. Each ciphertext $c_i$ is then split into $n$ shares, represented as $\{ s_1^i, s_2^i, \dots, s_n^i \}$, ensuring that no single party can reconstruct the plaintext.

Homomorphic encryption preserves the structure of operations such that the product of two ciphertexts results in the encryption of the product of their plaintexts. Specifically, for each data element $v_i$ and corresponding query element $q_i$, the homomorphic property ensures that $Enc(v_i, pk) \cdot Enc(q_i, pk) = Enc(v_i \cdot q_i, pk)$.

Each party $P_j$ receives a share $s_j^i$ and performs homomorphic operations with the corresponding encrypted query $Enc(q_i, pk)$. This operation produces a partial result $r_j^i = Enc(v_i \cdot q_i, pk)$ for each $i$, and the sum of these partial results is aggregated across all parties: $\sum_{j=1}^{n} r_j^i$. The final aggregated ciphertext $\text{result}$ thus encrypts the sum of all homomorphic operations, i.e., $\text{result} = Enc\left( \sum_{i=1}^{m} (v_i \cdot q_i), pk \right)$.

The querying party then decrypts the aggregated result using the private key $sk$, yielding $\mathbf{v}_{\text{result}} = Dec(\text{result}, sk) = \sum_{i=1}^{m} (v_i \cdot q_i)$. This final result matches the correct computation on the original plaintext data.

Thus, the correctness of the SK-MHE protocol is guaranteed because each homomorphic operation preserves the underlying computation, and the aggregation of shares accurately reflects the sum of operations on the original data.

\subsection{Security Analysis}
The security of the proposed SK-MHE is demonstrated by the following theorem.

\begin{theorem}
The SK-MHE protocol is IND-CPA secure, assuming that the underlying homomorphic encryption scheme is IND-CPA secure.
\end{theorem}

\begin{proof}
Assume there exists an adversary $\mathcal{A}$ that can distinguish between the encryptions of two chosen plaintexts $\mathbf{v}_0 = (v_{0,1}, v_{0,2}, \dots, v_{0,m})$ and $\mathbf{v}_1 = (v_{1,1}, v_{1,2}, \dots, v_{1,m})$ under the SK-MHE protocol with a probability of success $1/2 + \epsilon(n)$, where $\epsilon(n)$ is a non-negligible function of the security parameter $n$. This assumption implies a break of the underlying homomorphic encryption scheme's IND-CPA security.

A function $\epsilon(n)$ is defined as negligible if for every polynomial $p(n)$, there exists an integer $n_0$ such that $\forall n > n_0$, $\epsilon(n) < \frac{1}{p(n)}$. In other words, $\epsilon(n)$ decreases faster than the inverse of any polynomial in $n$. If $\mathcal{A}$ has a success probability $1/2 + \epsilon(n)$, where $\epsilon(n)$ is non-negligible, this implies that $\epsilon(n)$ does not meet the criteria for a negligible function.

The SK-MHE protocol splits the ciphertext $\mathbf{c}_b$ into $n$ shares for each ciphertext element: $\mathbf{c}_b = \{ s_1^i, s_2^i, \dots, s_n^i \}$ for $i \in [1, m]$. Each share $s_j^i$ is processed using homomorphic operations, and the partial results $r_j^i$ are aggregated as $r_j^i = HomomorphicOp(s_j^i, Enc(q_i, pk)) = Enc(v_b^i \cdot q_i, pk)$. The final result is obtained by aggregating the partial results: $\text{result} = \sum_{j=1}^{n} \sum_{i=1}^{m} r_j^i = Enc\left( \sum_{i=1}^{m} (v_b^i \cdot q_i), pk \right)$, and the querying party decrypts the final result as $\mathbf{v}_{\text{result}} = Dec(\text{result}, sk) = \sum_{i=1}^{m} (v_b^i \cdot q_i)$.

If $\mathcal{A}$ can distinguish between the encryptions of $\mathbf{v}_0$ and $\mathbf{v}_1$ with a probability $1/2 + \epsilon(n)$, a reduction $\mathcal{S}$ can be constructed to break the IND-CPA security of the underlying homomorphic encryption scheme. The reduction $\mathcal{S}$ interacts with the adversary $\mathcal{A}$ by forwarding the public key $pk$ from the IND-CPA challenger. After $\mathcal{A}$ selects two plaintext vectors $\mathbf{v}_0$ and $\mathbf{v}_1$, $\mathcal{S}$ submits them to the IND-CPA challenger, which returns a challenge ciphertext $\mathbf{c}_b$. $\mathcal{S}$ simulates the SK-MHE protocol by splitting the ciphertext $\mathbf{c}_b$ into shares and performing homomorphic operations. The result is passed to $\mathcal{A}$, which outputs a guess $b'$ for $b$. The reduction uses this guess to win the IND-CPA game.

Since $\mathcal{A}$ can guess $b$ with a probability of $1/2 + \epsilon(n)$, $\mathcal{S}$ has the same advantage in distinguishing the encryptions in the IND-CPA game. The success probability of $\mathcal{S}$ is $\Pr[\mathcal{S} \text{ wins}] = \frac{1}{2} + \epsilon(n)$.

The reduction runs in polynomial time with respect to the security parameter $n$. Each step in the SK-MHE protocol—ciphertext splitting, homomorphic operations, and aggregation—is performed in polynomial time. Thus, the time complexity of the reduction is $O(\text{poly}(n))$. Since the reduction is polynomial-time and $\epsilon(n)$ is non-negligible, $\mathcal{S}$ can break the IND-CPA security of the underlying homomorphic encryption scheme with a non-negligible advantage.

This leads to a contradiction, as the underlying homomorphic encryption scheme is assumed to be IND-CPA secure. Therefore, $\mathcal{A}$ cannot distinguish between the encryptions of $\mathbf{v}_0$ and $\mathbf{v}_1$ in the SK-MHE protocol with a probability better than random guessing, except with negligible probability.

Thus, the SK-MHE protocol is IND-CPA secure.
\end{proof}

\subsection{Complexity Analysis}

\paragraph{Time Complexity} The time complexity of the SK-MHE protocol can be broken down into four phases: encryption, ciphertext splitting, homomorphic operations, and decryption. In the encryption phase, given a data vector $\mathbf{v} = (v_1, v_2, \dots, v_m)$, encrypting each element takes $O(1)$ time, leading to a total time complexity of $T_{\text{enc}} = O(m)$. Each ciphertext $c_i$ is then split into $n$ shares, where each share is generated in constant time, resulting in a splitting complexity of $T_{\text{split}} = O(mn)$. In the homomorphic operations phase, each share $s_j^i$ undergoes a homomorphic operation with the corresponding encrypted query element $Enc(q_i, pk)$, which takes constant time, giving a total complexity of $T_{\text{hom}} = O(mn)$. Finally, the aggregation and decryption phases involve summing results from all parties and decrypting the aggregated ciphertext. Aggregation takes $O(n)$ time, while decryption takes $O(1)$, resulting in $T_{\text{agg+dec}} = O(n) + O(1)$. Therefore, the total time complexity of the protocol is $T_{\text{total}} = O(mn)$.

\paragraph{Network Complexity} The network complexity of the protocol is measured by the total number of messages exchanged and the number of communication rounds. For each ciphertext $c_i$, $n$ shares are distributed to $n$ parties, requiring $n$ messages for each of the $m$ elements, resulting in $M_{\text{send}} = O(mn)$ messages for the share distribution. After performing the homomorphic operations, each party sends its partial results back to the querying party, involving $M_{\text{recv}} = O(mn)$ messages. The total number of messages exchanged during the protocol is $M_{\text{total}} = O(mn) + O(mn) = O(mn)$. The protocol requires two communication rounds: one for distributing the shares and one for collecting the results, so the number of communication rounds is $R_{\text{total}} = 2$.

\section{Multiplicative Caching}
\label{sec:mc}

\subsection{Protocol Description}

The multiplicative caching protocol optimizes the homomorphic encryption of vectors by precomputing and caching scalar values used during query operations. Given a data vector $\mathbf{v} = (v_1, v_2, \dots, v_m)$ and a query vector $\mathbf{q} = (q_1, q_2, \dots, q_m)$, this protocol leverages homomorphic encryption and caching mechanisms to reduce computational complexity while maintaining security.

During the preprocessing phase, for each element $v_i$ in the vector $\mathbf{v}$, an encrypted scalar value $Enc(\Delta_i)$ is computed and cached. Here, $\Delta_i$ represents a precomputed multiplicative scaling factor for the element $v_i$, and the cached value is denoted as $\text{cache}[i] = Enc(\Delta_i)$. This ensures that, during the query phase, the homomorphic multiplication between $\mathbf{v}$ and $\mathbf{q}$ can be performed efficiently without redundant computation.

When the query vector $\mathbf{q}$ is received, the system retrieves the corresponding cached values and computes the homomorphic product for each element. The encrypted scalar product for each $i$ is computed as $r_i = Enc(v_i) \cdot \text{cache}[i] \cdot Enc(q_i)$. After computing the scalar products for all elements, the results are aggregated as $\text{result} = \sum_{i=1}^{m} r_i$. This final result remains encrypted.

Once the aggregation is complete, the querying party decrypts the final result using the private key $sk$, yielding the decrypted scalar product $\mathbf{v}_{\text{result}} = Dec(\text{result}, sk)$. This protocol significantly reduces the computational overhead of homomorphic multiplication and aggregation for vector data, especially in large-scale vector databases.

The multiplicative caching protocol is formally outlined in Algorithm~\ref{alg:mc}.

\begin{algorithm}[!t]
\SetAlgoLined
\caption{Multiplicative Caching Protocol}
\label{alg:mc}
\KwIn{Data vector $\mathbf{v} = (v_1, v_2, \dots, v_m)$, query vector $\mathbf{q} = (q_1, q_2, \dots, q_m)$, public key $pk$, private key $sk$, cached values $\text{cache}[i] = Enc(\Delta_i)$}
\KwOut{Decrypted scalar product $\mathbf{v}_{\text{result}}$}

\For{$i = 1 \dots m$}{
    $c_i \gets Enc(v_i, pk)$ \\
    $p_i \gets Enc(q_i, pk)$ \\
    $s_i \gets \text{cache}[i] \cdot p_i$
}

\For{$i = 1 \dots m$}{
    $r_i \gets c_i \cdot s_i$
}

\For{$i = 1 \dots m$}{
    \text{if} $r_i \text{ requires normalization}$: \\
    \[
    r_i \gets r_i \cdot Enc(\frac{1}{\Delta_i}, pk)
    \]
}

$\text{result}_i \gets \sum_{i=1}^{m} r_i$

$e_{\text{result}} \gets Enc(\text{result}_i, pk)$

\For{$i = 1 \dots m$}{
    $r_i^{\prime} \gets \frac{r_i}{e_{\text{result}}}$
}

$\mathbf{v}_{\text{result}} \gets Dec(\sum_{i=1}^{m} r_i^{\prime}, sk)$

\Return $\mathbf{v}_{\text{result}}$
\end{algorithm}

\subsection{Correctness Analysis}

The correctness of the Multiplicative Caching Protocol relies on the properties of homomorphic encryption and the correct aggregation of encrypted scalar products. Given an encrypted data vector $\mathbf{v} = (v_1, v_2, \dots, v_m)$ and an encrypted query vector $\mathbf{q} = (q_1, q_2, \dots, q_m)$, the protocol ensures that the decrypted result accurately represents the scalar product of the original plaintext vectors.

For each element $v_i$ in the data vector $\mathbf{v}$, the precomputed encrypted value $\text{cache}[i] = Enc(\Delta_i)$ is used to scale the homomorphic operation. During the query phase, the encrypted scalar product for each $i$ is computed as $r_i = Enc(v_i) \cdot \text{cache}[i] \cdot Enc(q_i)$, where $\Delta_i$ is a scaling factor that adjusts for the precision of $v_i$. Homomorphic encryption guarantees that the product of two encrypted values corresponds to the encrypted product of the original plaintexts, i.e.,
\[
Enc(v_i) \cdot Enc(q_i) = Enc(v_i \cdot q_i).
\]
Thus, the cached value $\text{cache}[i] = Enc(\Delta_i)$ ensures that the encrypted product $r_i$ is correctly scaled to handle floating-point operations.

The intermediate result $r_i = Enc(v_i \cdot q_i \cdot \Delta_i)$ ensures that both vectors $\mathbf{v}$ and $\mathbf{q}$ are homomorphically multiplied and appropriately scaled. The correctness of the homomorphic operations is preserved by the structure of the encryption scheme, which maintains the algebraic relationships between plaintext values. Specifically, for every $i$, the encryption of the scalar product $v_i \cdot q_i$ is modulated by the precomputed value $\Delta_i$, ensuring that the final result is accurate to the original vectors' precision.

Once all $r_i$ values have been computed, the protocol aggregates them into a single encrypted result:
\[
\text{result} = \sum_{i=1}^{m} r_i = Enc\left( \sum_{i=1}^{m} v_i \cdot q_i \cdot \Delta_i \right).
\]
This aggregation is performed homomorphically, preserving the integrity of the encrypted scalar products. Since each $r_i$ is an encrypted value, the aggregation results in an encrypted sum of all scalar products.

After aggregation, the querying party decrypts the final result using the private key $sk$. The decryption of the final result $\text{result}$ yields:
\[
\mathbf{v}_{\text{result}} = Dec\left( \sum_{i=1}^{m} Enc(v_i \cdot q_i \cdot \Delta_i), sk \right) = \sum_{i=1}^{m} v_i \cdot q_i \cdot \Delta_i.
\]
To recover the correct scalar product between the original vectors, the protocol performs a final normalization step, dividing by the scaling factors $\Delta_i$. Thus, the final result is given by:
\[
\mathbf{v}_{\text{correct}} = \sum_{i=1}^{m} \frac{v_i \cdot q_i \cdot \Delta_i}{\Delta_i} = \sum_{i=1}^{m} v_i \cdot q_i.
\]
This guarantees that the protocol returns the exact scalar product of the original plaintext vectors $\mathbf{v}$ and $\mathbf{q}$.

By leveraging the homomorphic properties of the encryption scheme and the caching of precomputed scaling factors, the Multiplicative Caching Protocol ensures that the encrypted operations and the final decrypted result are both mathematically correct and consistent with the expected scalar product. The use of precomputed values allows for efficient processing without sacrificing the accuracy of the final result.

\subsection{Security Analysis}

\begin{theorem}
The Multiplicative Caching Protocol is IND-CPA secure, assuming that the underlying homomorphic encryption scheme is IND-CPA secure.
\end{theorem}

\begin{proof}
We prove this by reduction. Assume there exists an adversary $\mathcal{A}$ capable of distinguishing the encryption of two chosen vectors $\mathbf{v}_0 = (v_{0,1}, v_{0,2}, \dots, v_{0,m})$ and $\mathbf{v}_1 = (v_{1,1}, v_{1,2}, \dots, v_{1,m})$ under the Multiplicative Caching Protocol with a probability greater than $\frac{1}{2} + \epsilon(n)$, where $\epsilon(n)$ is a non-negligible function of the security parameter $n$. We will show that such an adversary can be used to break the IND-CPA security of the underlying homomorphic encryption scheme.

Let $\mathcal{S}$ be a simulator that interacts with the IND-CPA challenger and uses the adversary $\mathcal{A}$ to break the encryption scheme. $\mathcal{S}$ forwards the public key $pk$ from the challenger to $\mathcal{A}$ and proceeds as follows.

For each vector $\mathbf{v}_0$ and $\mathbf{v}_1$ selected by $\mathcal{A}$, the simulator receives the ciphertext $\mathbf{c}_b = (c_{b,1}, c_{b,2}, \dots, c_{b,m})$ where $b \in \{0, 1\}$. For each element $v_{b,i}$ of $\mathbf{v}_b$, the ciphertext $c_{b,i} = Enc(v_{b,i}, pk)$ is computed by the IND-CPA challenger. During the preprocessing phase, the simulator generates the cached values $\text{cache}[i] = Enc(\Delta_i)$ for each $i \in [1, m]$.

The homomorphic operation for each $i$ is computed as $r_i = c_{b,i} \cdot \text{cache}[i] \cdot Enc(q_i, pk)$. By the homomorphic properties of the encryption scheme, this operation preserves the algebraic relationships between plaintexts:
\[
Enc(v_{b,i}) \cdot Enc(q_i) \cdot Enc(\Delta_i) = Enc(v_{b,i} \cdot q_i \cdot \Delta_i).
\]
Thus, the adversary $\mathcal{A}$ receives the set of encrypted scalar products $r_i$, and the aggregated result is computed as:
\[
\text{result} = \sum_{i=1}^{m} r_i = Enc\left( \sum_{i=1}^{m} v_{b,i} \cdot q_i \cdot \Delta_i \right).
\]
$\mathcal{A}$ attempts to distinguish whether the encrypted vector corresponds to $\mathbf{v}_0$ or $\mathbf{v}_1$ based on the result. If $\mathcal{A}$ can distinguish between $\mathbf{v}_0$ and $\mathbf{v}_1$ with a probability greater than $\frac{1}{2} + \epsilon(n)$, then $\mathcal{S}$ can use this information to break the IND-CPA security of the encryption scheme.

We now show that the reduction $\mathcal{S}$ is efficient and operates in polynomial time with respect to the security parameter $n$. Each encryption $Enc(v_i, pk)$, homomorphic operation, and aggregation step is performed in $O(1)$ time for each element, resulting in a total complexity of $O(m)$ for $m$ vector elements. Thus, the reduction $\mathcal{S}$ runs in $O(m)$ time, which is polynomial in the size of the input.

The success probability of $\mathcal{S}$ is the same as that of $\mathcal{A}$, i.e., $\frac{1}{2} + \epsilon(n)$. Since $\mathcal{A}$ can distinguish between the encryptions of $\mathbf{v}_0$ and $\mathbf{v}_1$ with non-negligible probability $\epsilon(n)$, this implies that $\mathcal{S}$ can break the IND-CPA security of the homomorphic encryption scheme with the same advantage.

This leads to a contradiction, as the homomorphic encryption scheme is assumed to be IND-CPA secure. Therefore, no polynomial-time adversary can distinguish between the encryptions of two vectors $\mathbf{v}_0$ and $\mathbf{v}_1$ with a probability better than random guessing, except with negligible probability. Hence, the Multiplicative Caching Protocol is IND-CPA secure.
\end{proof}

\subsection{Complexity Analysis}

\paragraph{Time Complexity} The time complexity of the Multiplicative Caching Protocol consists of three phases: encryption, homomorphic operations, and aggregation. For a data vector $\mathbf{v} = (v_1, v_2, \dots, v_m)$ and a query vector $\mathbf{q} = (q_1, q_2, \dots, q_m)$, each element in $\mathbf{v}$ and $\mathbf{q}$ is encrypted in constant time $O(1)$, giving a total encryption time of $T_{\text{enc}} = O(m)$. Homomorphic operations, including multiplying the cached value $\text{cache}[i]$ with encrypted data and query elements, also take $O(1)$ time per element, resulting in $T_{\text{hom}} = O(m)$. Finally, aggregation of $m$ encrypted scalar products requires $O(m)$ time. Thus, the total time complexity is $T_{\text{total}} = O(m)$.

\paragraph{Network Complexity} The network complexity is determined by the number of messages exchanged and communication rounds. For both $\mathbf{v}$ and $\mathbf{q}$, $m$ encrypted elements are sent to the server, resulting in $M_{\text{enc}} = O(m)$ messages. After the homomorphic operations, $m$ encrypted results are sent back, with $M_{\text{recv}} = O(m)$ messages. The total number of messages exchanged is $M_{\text{total}} = O(m)$, and the protocol requires two communication rounds, $R_{\text{total}} = 2$.

\section{System Implementation}

\subsection{SK-MHE Implementation}

The SK-MHE protocol is the backbone of secure computations in our system, enabling encrypted Approximate $k$-Nearest Neighbor (ANN) searches across distributed, mutually-distrusted nodes. Our implementation of SK-MHE is built on top of the FedML framework~\cite{fedml}, which supports federated learning and distributed machine learning workflows. FedML allows seamless integration of the message-passing interface (MPI)~\cite{openmpi} to enable communication between nodes in a geographically distributed setup.

The FedML framework was chosen for its flexible support of large-scale federated architectures. We leverage the MPI interface for efficient inter-node communication, using a unified namespace (\textit{MPI\_COMM\_WORLD}) to manage communication. Each node is assigned a single CPU core (referred to as a \textit{rank} in MPI), optimizing resource allocation and minimizing communication overhead. This allows the system to scale across geographically distributed data centers, while ensuring that all nodes can securely participate in encrypted computations.

For homomorphic encryption operations, we use the TenSEAL library~\cite{tenseal}, which is a popular Python library for homomorphic encryption on tensors. Specifically, we employ the CKKS encryption scheme~\cite{ckks}, which is ideal for working with high-dimensional floating-point data, such as vector embeddings used in ANN searches. The CKKS scheme allows secure arithmetic operations on encrypted vectors, such as dot product calculations and distance measurements, without revealing the underlying data. 

The implementation of SK-MHE involves creating encrypted query vectors and performing secure vector computations across distributed nodes. The core homomorphic operations, such as encrypted vector multiplication (\textit{EvalMul}) and addition (\textit{EvalAdd}), are implemented using TenSEAL’s APIs. We have extended the FedML architecture to support secure aggregation and decryption steps after each computation, ensuring that no single node has access to decrypted results. Instead, only the query owner can decrypt the final aggregated result using their private key.

The integration of TenSEAL into FedML required adjustments to the communication pipeline. Since homomorphic operations are computationally intensive, we utilized asynchronous MPI communications to prevent blocking during encryption, multiplication, and addition operations. This asynchronous design reduces latency and ensures that the system remains responsive even when processing large encrypted datasets.

\subsection{Multiplicative Caching Implementation}

The Multiplicative Caching (MC) protocol is implemented to optimize the performance of homomorphic computations in the FRAG system. This protocol precomputes and caches encrypted scalar products to avoid redundant calculations during encrypted ANN searches. Our MC implementation uses Redis~\cite{redis}, a distributed in-memory data store, to handle the caching of encrypted scalar values. Redis is chosen for its ability to handle high-throughput data retrieval with low latency, making it ideal for a federated setup where multiple nodes need rapid access to precomputed values.

To implement MC, we designed a caching layer in Redis that stores encrypted scalar products for each vector element. These cached values are indexed by vector IDs and can be retrieved during query processing to perform secure multiplications without recomputing the encrypted values. By precomputing these values during the offline phase, we minimize the computational cost during the online query phase. Each node has access to a Redis instance, and we leverage Redis clustering to ensure that cached values are available globally across all participating nodes.

We integrated the caching mechanism with our SK-MHE implementation by modifying the FedML communication pipeline to include Redis queries. When a query vector is received by a node, the system checks Redis for precomputed encrypted values corresponding to the query vector's dimensions. If a cached value is found, it is retrieved and used in the homomorphic computation, bypassing the need for real-time encryption. This caching mechanism significantly reduces the time required to process each query, especially when large datasets are involved.

Our implementation also includes a cache invalidation mechanism to ensure that cached values remain up to date. This mechanism is crucial in dynamic environments where vector embeddings are frequently updated. Redis’ native support for TTL (time-to-live) values ensures that stale cache entries are automatically removed after a configurable time period, further optimizing the system's performance.

\subsection{Integration into FRAG}

Integrating SK-MHE and MC into the distributed Retrieval-Augmented Generation (RAG) system involves combining these protocols with the RAG's core functionality of generating relevant responses based on retrieved knowledge. Our implementation integrates these cryptographic protocols into a distributed ANN search pipeline, where encrypted query vectors are securely matched against distributed encrypted datasets without exposing sensitive information to any participating node.

The overall architecture is designed around FedML’s federated infrastructure, with each node hosting a subset of the vector database and participating in the encrypted ANN search process. Nodes exchange encrypted query and response vectors using MPI, and all homomorphic operations (both SK-MHE and MC) are performed locally at each node. This setup ensures that no plaintext vectors are ever exposed during the search process.

To integrate SK-MHE and MC, we designed a custom query processing pipeline. When a query is received, the querying node generates an encrypted query vector using the CKKS scheme. This vector is then broadcast to all participating nodes using the MPI interface. Each node retrieves the encrypted vector, performs the necessary homomorphic operations (using both SK-MHE for secure multiplication and MC for optimized caching), and sends back an encrypted result. These encrypted results are then aggregated by the querying node, and the final result is decrypted to reveal the $k$ nearest neighbors.

The distributed nature of the system ensures that nodes can collaborate on encrypted ANN searches without ever revealing their individual datasets or query vectors. The use of Redis for caching, combined with the efficiency of SK-MHE, allows the system to scale to large datasets and handle real-time queries efficiently. The integration of both protocols into the RAG system enables secure, federated knowledge retrieval in environments where data privacy is paramount.

Our final system is designed to be modular and scalable, with each protocol (SK-MHE and MC) functioning as independent components that can be optimized or replaced as needed. We plan to open-source our implementation as a Python package, integrated with FedML and TenSEAL, to facilitate the adoption of secure, federated RAG systems in a variety of applications.

\section{Evaluation}

\subsection{Experimental Setup}

To evaluate the performance and security of the FRAG system, we conducted a series of experiments on a distributed setup comprising multiple geographically-distributed nodes. Each node operates a local vector database (VecDB) and participates in the secure computation protocols, including Single-Key Multiparty Homomorphic Encryption (SK-MHE) and Multiplicative Caching (MC). The goal of the experiments is to measure the system’s efficiency, scalability, and security in a real-world federated environment, processing large datasets across distributed nodes.

\paragraph{Infrastructure and Environment}
The experiments were performed on a distributed infrastructure using 10 physical machines. Each machine is equipped with 4 Intel Xeon vCPUs, 192 GB RAM, and 2 TB SSD storage, running Ubuntu 22.04. The machines are connected via a high-bandwidth, low-latency network, representative of real-world distributed environments. Docker was used to containerize the FRAG system components, and Kubernetes was employed for orchestration and scalability management, enabling the dynamic addition and removal of nodes during the experiments.

\paragraph{Vector Databases (VecDBs)}
Each machine runs an instance of FAISS (Facebook AI Similarity Search) for Approximate k-Nearest Neighbor (ANN) searches. We use high-dimensional vector embeddings generated from real-world datasets, including OpenAI’s GPT-3 embeddings. The dataset consists of 1 million vectors, with each vector containing 768 dimensions. These embeddings are encrypted using the SK-MHE protocol before being stored in VecDBs, ensuring both data privacy and secure retrieval across distributed nodes.

\paragraph{Homomorphic Encryption and Multiplicative Caching}
The homomorphic encryption operations are implemented using the Microsoft SEAL library, with the CKKS scheme used for its ability to handle floating-point computations. SK-MHE is employed to securely perform homomorphic operations on encrypted vectors across the distributed nodes, such as dot product calculations and distance measurements for ANN searches. The Multiplicative Caching protocol is used to optimize these homomorphic computations by caching frequently-used encrypted scalar products. Redis is used as the distributed caching layer, providing fast access to precomputed encrypted values.

\paragraph{Performance Metrics}
The performance of the FRAG system was evaluated across three key metrics: query response time, computational overhead, and memory consumption. Query response time is defined as the time between the submission of a query and the retrieval of the final result, including all encryption, ANN search, and decryption steps. Computational overhead is measured by tracking the CPU utilization and the number of homomorphic operations executed on each node. Memory consumption is analyzed by monitoring the usage of Redis as a cache layer for storing intermediate encrypted values and comparing the resource usage with and without caching.

\paragraph{Experiment Execution}
We executed a series of ANN search queries on the distributed VecDBs, varying the size of the query vectors and the number of participating nodes. The number of nodes ranged from 3 to 10, and we tested with query vector datasets ranging from 100,000 to 1 million vectors. Each query was executed 10 times to ensure consistency in results. The experiments were conducted under normal network conditions to simulate the type of load and latency expected in a real-world federated system.

\subsection{SK-MHE Performance}

The objective of this experiment is to evaluate the performance and scalability of the Single-Key Multiparty Homomorphic Encryption (SK-MHE) protocol within the FRAG system. We assess SK-MHE based on two key metrics: the performance of cryptographic primitives (Split, Merge, Verify), and the overhead introduced by SK-MHE compared to a baseline (FedAvg). These experiments were conducted on various datasets, including MNIST, FMNIST, CIFAR-10, and SVHN, to ensure robustness across different scenarios. All time measurements in this section are reported in seconds, unless otherwise stated.

\begin{figure}[t]
    \centering
    \includegraphics[width=85mm]
    {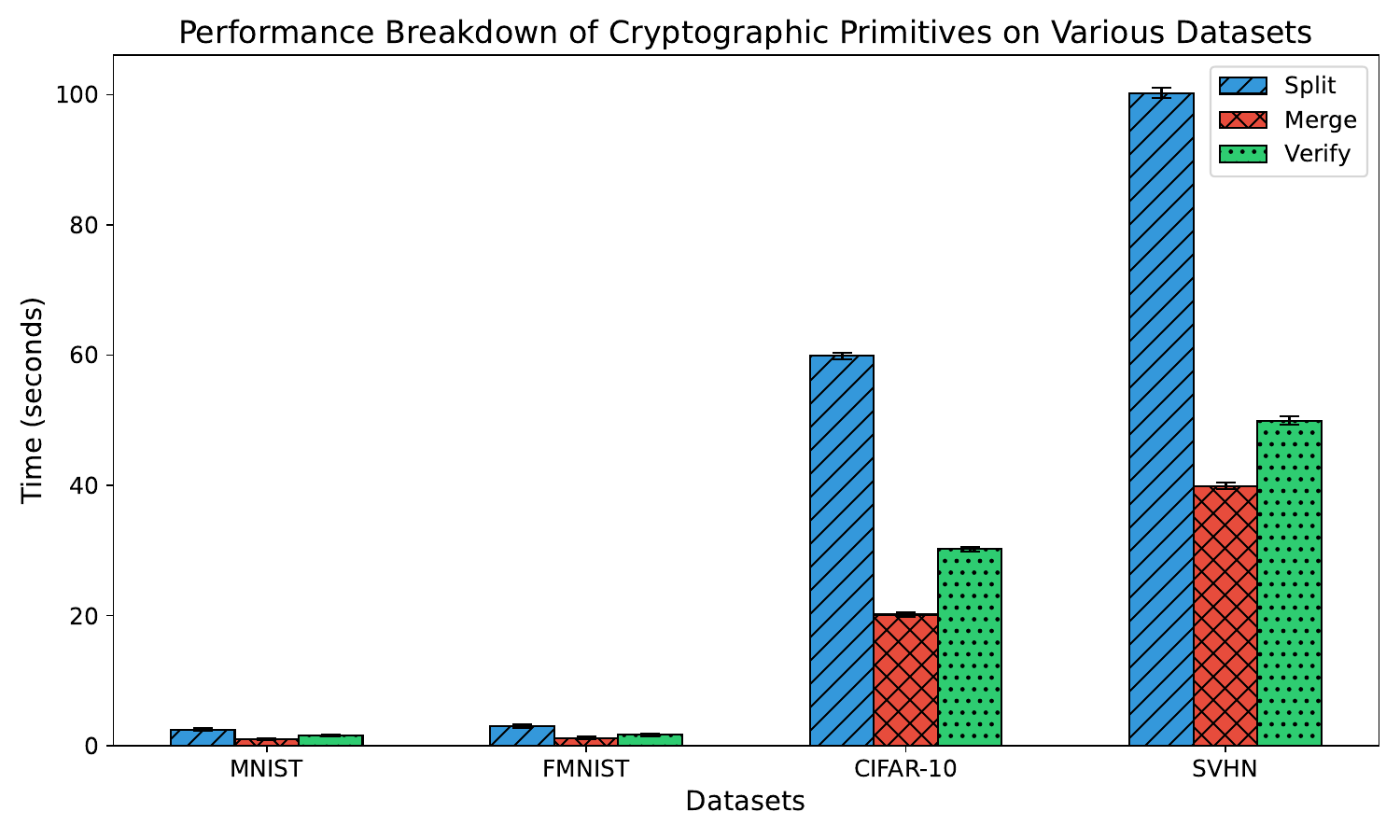}
    \caption{Performance Breakdown of cryptographic primitives on MNIST, FMNIST, CIFAR-10, and SVHN.}
    \label{fig:atss}
\end{figure}

\paragraph{Performance of Cryptographic Primitives}
We measure the time taken for three core cryptographic primitives (Split, Merge, Verify) that are integral to the SK-MHE protocol. Figure~\ref{fig:atss} shows the performance of these primitives on MNIST, FMNIST, CIFAR-10, and SVHN. As expected, SK-MHE performs efficiently on simpler datasets such as MNIST and FMNIST. However, larger datasets like CIFAR-10 and SVHN exhibit longer times due to the increased complexity of data encryption and verification processes.

\begin{figure}[t]
    \centering
    \includegraphics[width=85mm]
    {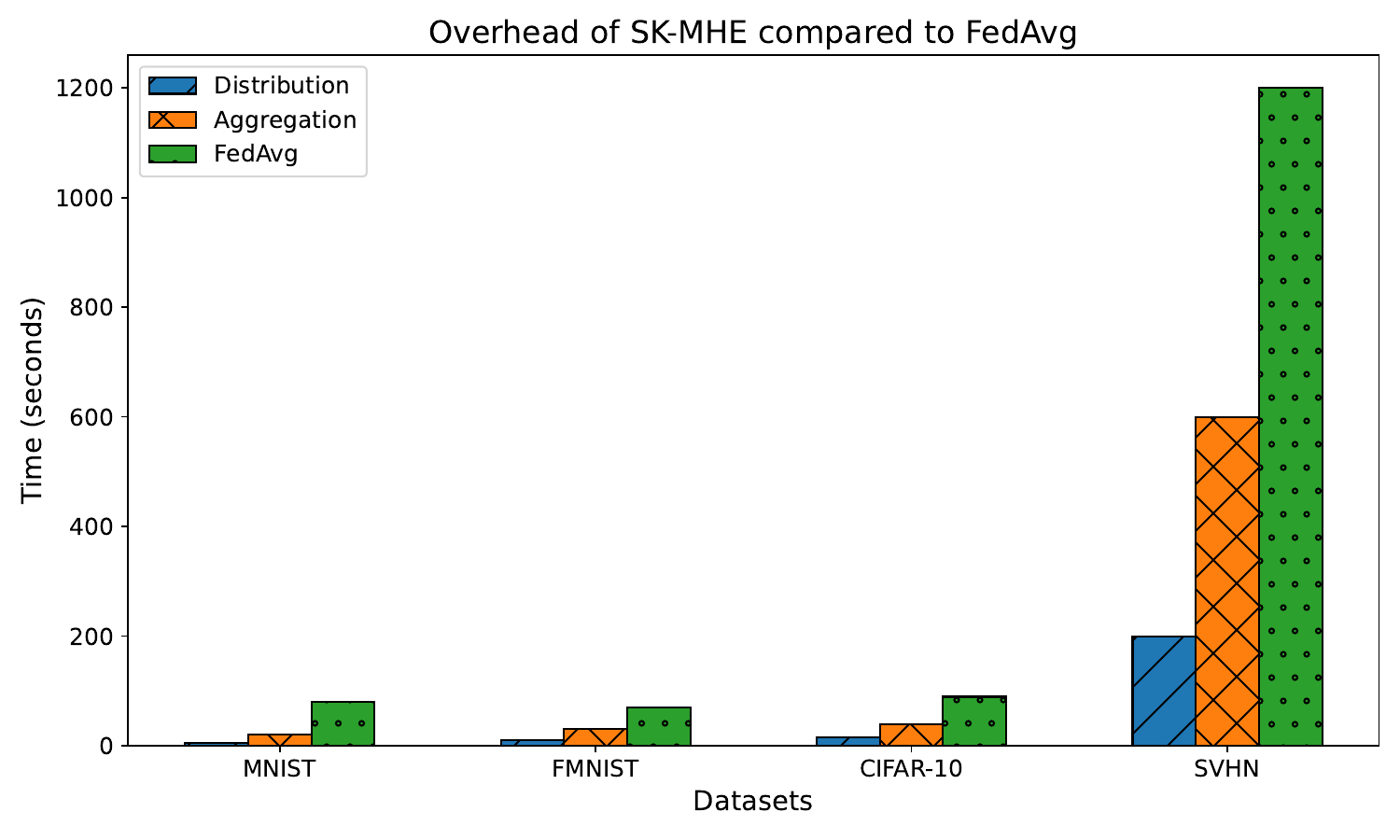}
    \caption{Overhead of SK-MHE compared to FedAvg on MNIST, FMNIST, CIFAR-10, and SVHN (time in seconds).}
    \label{fig:overhead}
\end{figure}

\paragraph{Overhead Evaluation}
Next, we compare the overhead introduced by SK-MHE against a non-encrypted federated protocol (FedAvg). Figure~\ref{fig:overhead} presents the time spent on distribution, aggregation, and the FedAvg baseline for each dataset. As expected, the secure operations in SK-MHE introduce overhead, particularly in aggregation steps, but the overall performance remains competitive. The results suggest that SK-MHE is well-suited for federated environments where both security and efficiency are paramount.

\subsection{Performance of Multiplicative Caching (MC)}

\paragraph{Computational Cost of Multiplicative Caching}

\begin{figure}[t]
    \centering
    \includegraphics[width=85mm]
    {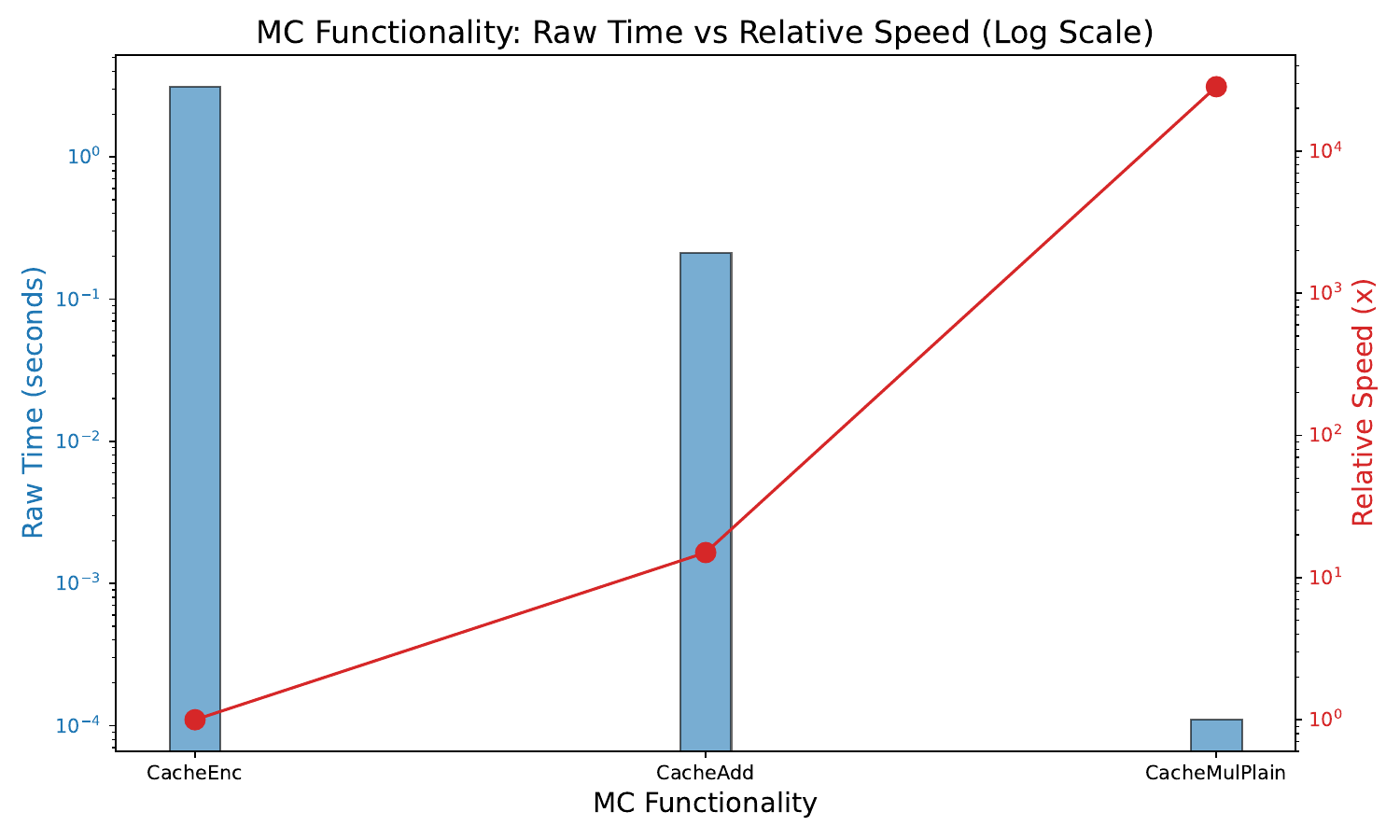}
    \caption{Computational cost of multiplicative caching algorithms in MySQL loadable functions.}
    \label{fig:func_mc}
\end{figure}

We evaluate the performance of multiplicative caching integrated with MySQL loadable functions, focusing on three operations: multiplication between a plaintext and a cached encrypted value (\textit{CacheMulPlain}), addition between two cached encrypted values (\textit{CacheAdd}), and caching of a plaintext (\textit{CacheEnc}). Figure~\ref{fig:func_mc} shows the results for 1,000 repetitions with random floating-point numbers between 1 and 20,000. The results indicate that addition between two cached encrypted values is significantly faster than caching itself, while plaintext-cached encrypted multiplication is several orders of magnitude faster than caching.

\paragraph{Performance Overhead of Multiplicative Caching}

\begin{figure}[t]
    \centering
    \includegraphics[width=85mm]
    {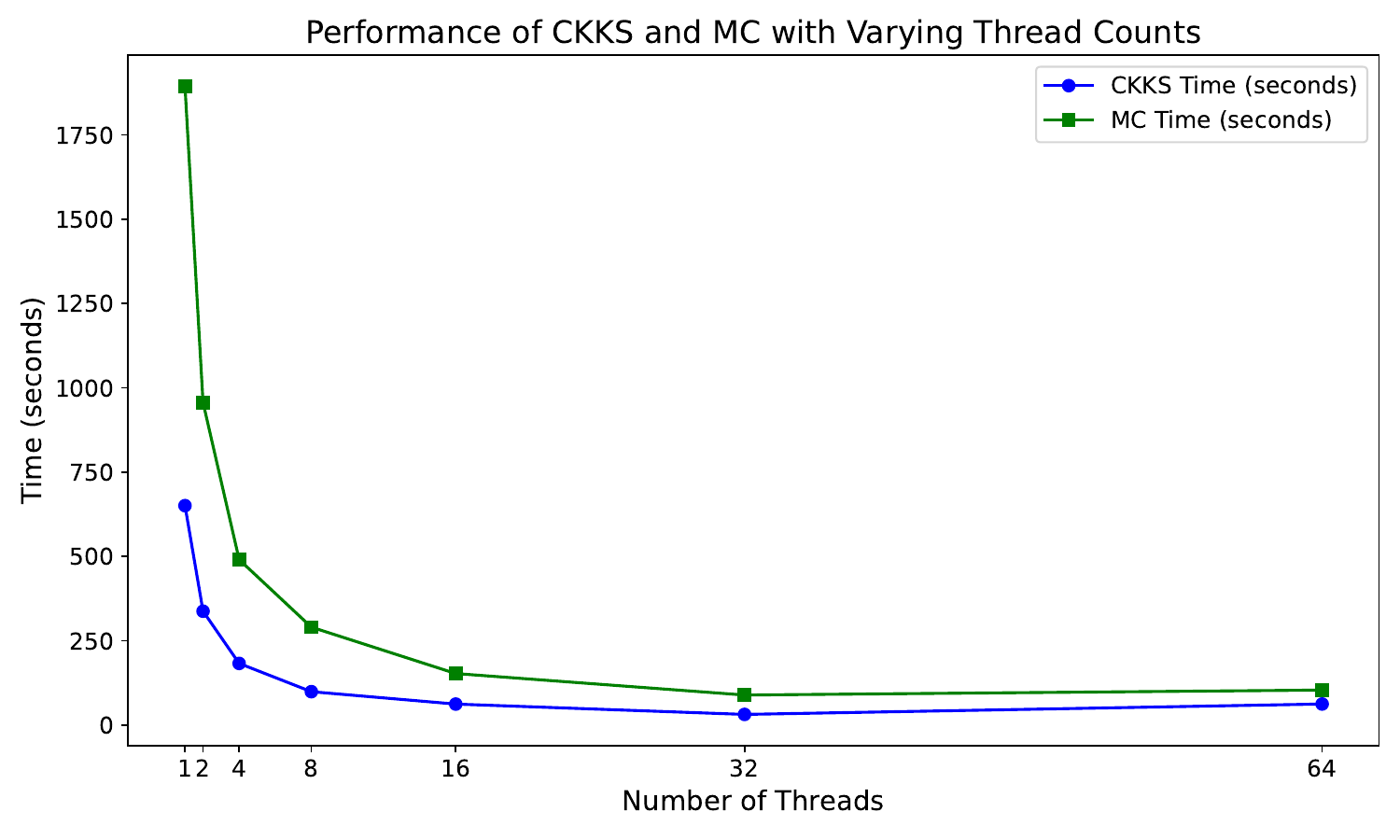}
    \caption{Cost of different thread counts for caching ciphertexts.}
    \label{fig:openmp_mc}
\end{figure}

We measured the performance overhead introduced by the Multiplicative Caching protocol by evaluating both time and memory usage during the caching of 200,000 plaintexts. The evaluation was conducted using varying numbers of threads, ranging from 1 to 96, on a system equipped with dual CPUs. The results, as shown in Figure~\ref{fig:openmp_mc}, indicate that while the caching process scales effectively with the increase in thread count, performance improvements begin to plateau beyond 32 threads. This performance degradation is primarily attributed to the overhead introduced by context switching, which becomes increasingly pronounced when using more threads.

\paragraph{Scalability of Multiplicative Caching}

\begin{figure}[t]
    \centering
    \includegraphics[width=85mm]
    {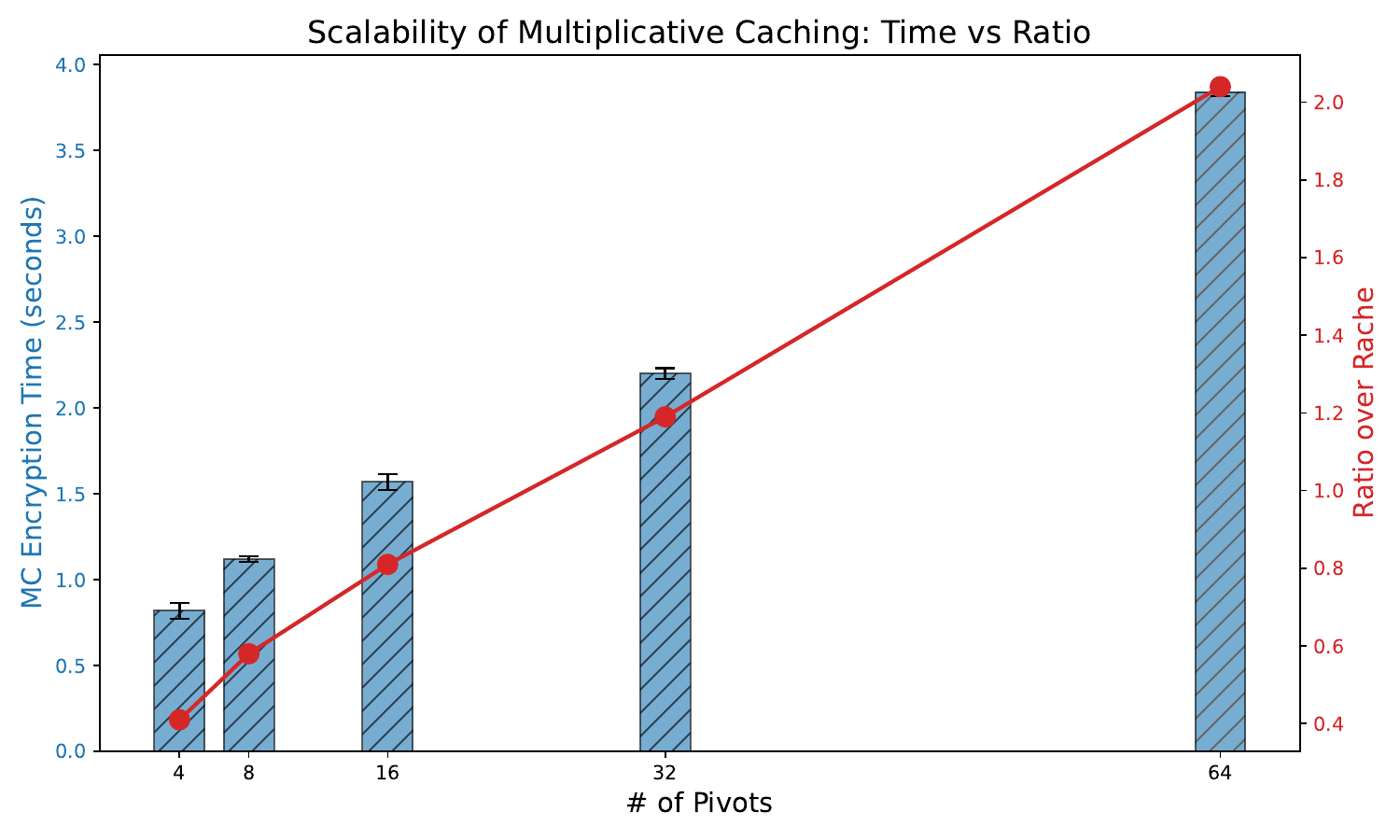}
    \caption{Scalability of Multiplicative Caching with different pivots}
    \label{fig:mc_scale}
\end{figure}

We explored the scalability of the Multiplicative Caching (MC) protocol by analyzing how the computational cost evolves as the number of precomputed encrypted pivots increases. The primary focus of this experiment was to determine how the system's performance scales with the number of cached values used in the Approximate $k$-Nearest Neighbor (ANN) searches. As shown in Figure~\ref{fig:mc_scale}, the computational time increases as the number of pivots grows from 4 to 64, revealing key insights into the trade-offs between caching efficiency and computational overhead.

\paragraph{Computational Overhead of MC}

\begin{figure}[t]
    \centering
    \includegraphics[width=85mm]
    {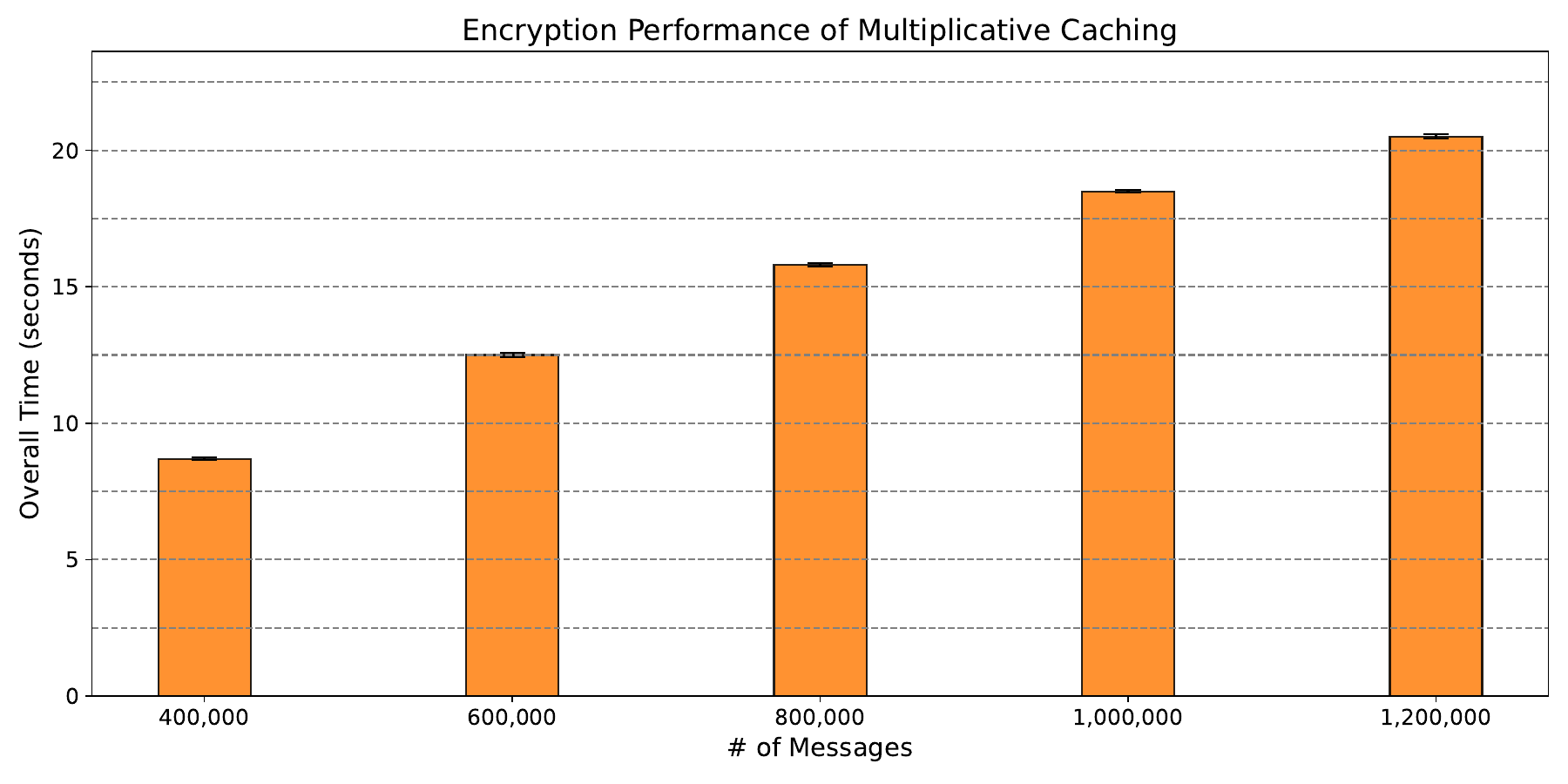}
    \caption{Encryption performance of Multiplicative Caching}
    \label{fig:mc_scale2}
\end{figure}

To assess the computational overhead of the Multiplicative Caching (MC) protocol, we conducted an experiment where the number of plaintext messages processed varied from 40 to 120. The goal of this evaluation was to observe the weak-scaling behavior of the MC protocol, where the focus is on whether the computational time scales proportionally to the number of messages. As depicted in Figure~\ref{fig:mc_scale2}, the time required per message remains relatively constant, even as the number of messages increases. This consistent per-message time suggests that the overhead introduced by MC is primarily independent of the number of plaintext messages being processed.

\paragraph{Real-world Data Sets}

\begin{figure}[t]
    \centering
    \includegraphics[width=85mm]
    {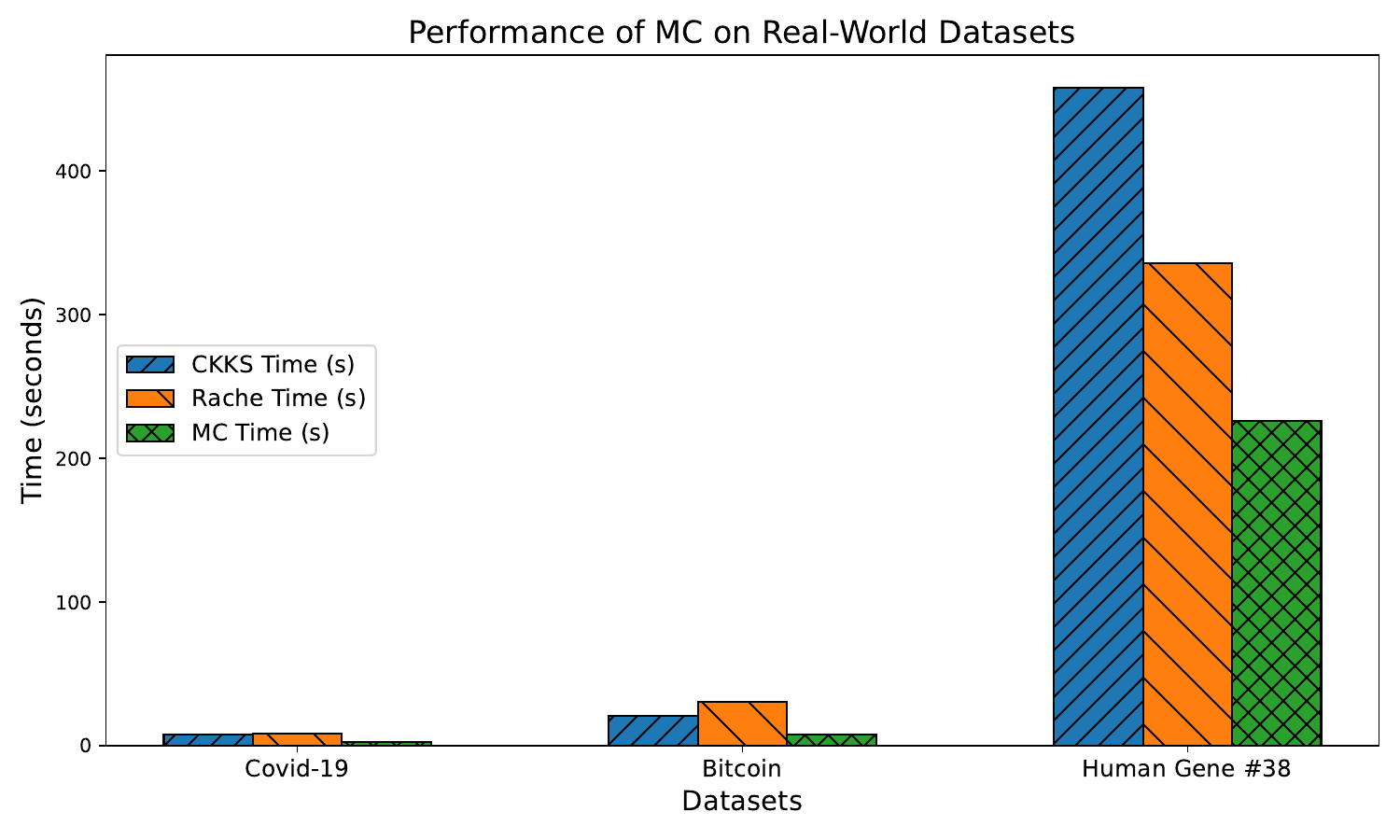}
    \caption{Performance comparison of MC with real-world data sets}
    \label{fig:mc_apps}
\end{figure}

To evaluate the effectiveness and real-world applicability of the Multiplicative Caching (MC) protocol, we tested its performance across three diverse and large-scale datasets: Covid-19, Bitcoin, and Human Gene \#38. Each dataset poses unique challenges in terms of size, complexity, and data retrieval patterns, providing a comprehensive benchmark for assessing MC's efficiency in practical scenarios. As shown in Figure~\ref{fig:mc_apps}, MC consistently outperformed the CKKS scheme and non-caching approaches, achieving speedups ranging from 1.46$\times$ to 2.61$\times$, depending on the dataset.

For the Covid-19 dataset, which primarily includes time-series data, the system demonstrated a significant speedup of 2.61$\times$ when using MC compared to CKKS. The large number of queries associated with the evolving nature of the dataset makes it an ideal candidate for testing MC's ability to handle dynamic, frequently updated data. By caching frequently used encrypted scalar values, the overhead of recomputing these values was drastically reduced, resulting in faster query responses and improved system efficiency.

In the Bitcoin dataset, which is characterized by high transaction volume and frequent updates, MC also provided a considerable speedup of 2.60$\times$. The ability to handle such a high-volume, high-throughput dataset without introducing significant latency showcases MC's suitability for real-time applications where fast query responses are essential. The dataset’s size and complexity highlight the necessity of efficient caching mechanisms, as recomputing scalar products on encrypted vectors could otherwise cause substantial delays. MC's ability to optimize these computations significantly reduced the overall query processing time.

The Human Gene \#38 dataset, which consists of a large collection of genomic sequences, presents one of the most computationally demanding challenges due to its vast size and the complexity of the operations required for ANN searches. Even in this highly demanding scenario, MC exhibited a notable speedup of 1.46$\times$. Although the improvement here is slightly lower compared to the other datasets, the results still underscore MC’s capacity to handle computationally intensive tasks in large-scale databases. The high dimensionality and intricate relationships within the genomic data required complex calculations, yet MC effectively optimized these operations by reducing the redundant recomputation of scalar values, maintaining a balance between computational efficiency and accuracy.

\section{Conclusion}

In this paper, we presented \textit{FRAG}, a Federated Retrieval-Augmented Generation system that enables secure and efficient distributed approximate $k$-nearest neighbor (ANN) searches over encrypted vector databases. FRAG addresses the challenges inherent in federated environments where mutually distrusted parties must collaborate, without exposing sensitive data. The system leverages two innovative protocols: the Single-Key Homomorphic Encryption (SK-MHE) protocol and the Multiplicative Caching (MC) protocol.
The SK-MHE protocol ensures that secure computations can be performed over encrypted vectors, while simplifying key management, making it ideal for collaborative environments that prioritize data privacy. Meanwhile, the Multiplicative Caching protocol enhances performance by precomputing and caching frequently used scalar values, significantly reducing computation times, particularly for large-scale real-world datasets. 
Through comprehensive experimentation, we demonstrated the practical feasibility of FRAG on both synthetic and real-world datasets, including Covid-19, Bitcoin, and Human Gene \#38. 

\section{Acknowledgement}
Results presented in this paper were partly obtained using the Chameleon testbed supported by the National Science Foundation.

\bibliographystyle{ACM-Reference-Format}
\bibliography{sample-base}


\begin{thebibliography}{55}


\ifx \showCODEN    \undefined \def \showCODEN     #1{\unskip}     \fi
\ifx \showDOI      \undefined \def \showDOI       #1{#1}\fi
\ifx \showISBNx    \undefined \def \showISBNx     #1{\unskip}     \fi
\ifx \showISBNxiii \undefined \def \showISBNxiii  #1{\unskip}     \fi
\ifx \showISSN     \undefined \def \showISSN      #1{\unskip}     \fi
\ifx \showLCCN     \undefined \def \showLCCN      #1{\unskip}     \fi
\ifx \shownote     \undefined \def \shownote      #1{#1}          \fi
\ifx \showarticletitle \undefined \def \showarticletitle #1{#1}   \fi
\ifx \showURL      \undefined \def \showURL       {\relax}        \fi
\providecommand\bibfield[2]{#2}
\providecommand\bibinfo[2]{#2}
\providecommand\natexlab[1]{#1}
\providecommand\showeprint[2][]{arXiv:#2}

\bibitem[Alfeld et~al\mbox{.}(2016)]%
        {alfeld2016}
\bibfield{author}{\bibinfo{person}{Scott Alfeld}, \bibinfo{person}{Xiaojin Zhu}, {and} \bibinfo{person}{Paul Barford}.} \bibinfo{year}{2016}\natexlab{}.
\newblock \showarticletitle{Data poisoning attacks against autoregressive models}. In \bibinfo{booktitle}{\emph{Proceedings of the AAAI Conference on Artificial Intelligence}}.
\newblock


\bibitem[Bagdasaryan et~al\mbox{.}(2020)]%
        {BagdasaryanVHES20}
\bibfield{author}{\bibinfo{person}{Eugene Bagdasaryan}, \bibinfo{person}{Andreas Veit}, \bibinfo{person}{Yiqing Hua}, \bibinfo{person}{Deborah Estrin}, {and} \bibinfo{person}{Vitaly Shmatikov}.} \bibinfo{year}{2020}\natexlab{}.
\newblock \showarticletitle{How To Backdoor Federated Learning}. In \bibinfo{booktitle}{\emph{The 23rd International Conference on Artificial Intelligence and Statistics (AIStat)}}, \bibfield{editor}{\bibinfo{person}{Silvia Chiappa} {and} \bibinfo{person}{Roberto Calandra}} (Eds.).
\newblock


\bibitem[Benaissa et~al\mbox{.}(2021)]%
        {tenseal}
\bibfield{author}{\bibinfo{person}{Ayoub Benaissa}, \bibinfo{person}{Bilal Retiat}, \bibinfo{person}{Bogdan Cebere}, {and} \bibinfo{person}{Alaa~Eddine Belfedhal}.} \bibinfo{year}{2021}\natexlab{}.
\newblock \bibinfo{title}{TenSEAL: A Library for Encrypted Tensor Operations Using Homomorphic Encryption}.
\newblock
\newblock
\showeprint[arxiv]{2104.03152}~[cs.CR]


\bibitem[Bernardini and colleagues(2017)]%
        {bernardini2017annoy}
\bibfield{author}{\bibinfo{person}{Michael Bernardini} {and} \bibinfo{person}{colleagues}.} \bibinfo{year}{2017}\natexlab{}.
\newblock \showarticletitle{Annoy: Approximate nearest neighbors in C++/Python}.
\newblock \bibinfo{journal}{\emph{GitHub repository}} (\bibinfo{year}{2017}).
\newblock


\bibitem[Bhagoji et~al\mbox{.}(2019)]%
        {bhagoji2019analyzing}
\bibfield{author}{\bibinfo{person}{Arjun~Nitin Bhagoji}, \bibinfo{person}{Supriyo Chakraborty}, \bibinfo{person}{Prateek Mittal}, {and} \bibinfo{person}{Seraphin Calo}.} \bibinfo{year}{2019}\natexlab{}.
\newblock \showarticletitle{Analyzing federated learning through an adversarial lens}. In \bibinfo{booktitle}{\emph{International Conference on Machine Learning}}.
\newblock


\bibitem[Bonawitz et~al\mbox{.}(2017)]%
        {bonawitz2017practical}
\bibfield{author}{\bibinfo{person}{Keith Bonawitz}, \bibinfo{person}{Vladimir Ivanov}, \bibinfo{person}{Benjamin Kreuter}, \bibinfo{person}{Antonio Marcedone}, \bibinfo{person}{H~Brendan McMahan}, \bibinfo{person}{Sarvar Patel}, \bibinfo{person}{Daniel Ramage}, \bibinfo{person}{Aaron Segal}, {and} \bibinfo{person}{Karn Seth}.} \bibinfo{year}{2017}\natexlab{}.
\newblock \showarticletitle{Practical secure aggregation for privacy-preserving machine learning}. In \bibinfo{booktitle}{\emph{Proceedings of the 2017 ACM SIGSAC Conference on Computer and Communications Security}}. \bibinfo{pages}{1175--1191}.
\newblock


\bibitem[Brown et~al\mbox{.}(2020)]%
        {brown2020language}
\bibfield{author}{\bibinfo{person}{Tom Brown}, \bibinfo{person}{Benjamin Mann}, \bibinfo{person}{Nick Ryder}, \bibinfo{person}{Melanie Subbiah}, \bibinfo{person}{Jared Kaplan}, \bibinfo{person}{Prafulla Dhariwal}, \bibinfo{person}{Arvind Neelakantan}, \bibinfo{person}{Pranav Shyam}, \bibinfo{person}{Girish Sastry}, \bibinfo{person}{Amanda Askell}, {et~al\mbox{.}}} \bibinfo{year}{2020}\natexlab{}.
\newblock \showarticletitle{Language models are few-shot learners}.
\newblock \bibinfo{journal}{\emph{Advances in Neural Information Processing Systems}}  \bibinfo{volume}{33} (\bibinfo{year}{2020}), \bibinfo{pages}{1877--1901}.
\newblock


\bibitem[Cheon et~al\mbox{.}(2017)]%
        {ckks}
\bibfield{author}{\bibinfo{person}{Jung~Hee Cheon}, \bibinfo{person}{Andrey Kim}, \bibinfo{person}{Miran Kim}, {and} \bibinfo{person}{Yong~Soo Song}.} \bibinfo{year}{2017}\natexlab{}.
\newblock \showarticletitle{Homomorphic Encryption for Arithmetic of Approximate Numbers}. In \bibinfo{booktitle}{\emph{23rd International Conference on the Theory and Applications of Cryptology and Information Security (AsiaCrypt)}}, \bibfield{editor}{\bibinfo{person}{Tsuyoshi Takagi} {and} \bibinfo{person}{Thomas Peyrin}} (Eds.). \bibinfo{publisher}{Springer}.
\newblock


\bibitem[Data and Diggavi(2021)]%
        {ddata_icml21}
\bibfield{author}{\bibinfo{person}{Deepesh Data} {and} \bibinfo{person}{Suhas Diggavi}.} \bibinfo{year}{2021}\natexlab{}.
\newblock \showarticletitle{Byzantine-Resilient High-Dimensional SGD with Local Iterations on Heterogeneous Data}. In \bibinfo{booktitle}{\emph{Proceedings of the 38th International Conference on Machine Learning}} \emph{(\bibinfo{series}{Proceedings of Machine Learning Research}, Vol.~\bibinfo{volume}{139})}, \bibfield{editor}{\bibinfo{person}{Marina Meila} {and} \bibinfo{person}{Tong Zhang}} (Eds.). \bibinfo{publisher}{PMLR}, \bibinfo{pages}{2478--2488}.
\newblock
\urldef\tempurl%
\url{https://proceedings.mlr.press/v139/data21a.html}
\showURL{%
\tempurl}


\bibitem[Demmler et~al\mbox{.}(2015)]%
        {aby}
\bibfield{author}{\bibinfo{person}{Daniel Demmler}, \bibinfo{person}{Thomas Schneider}, {and} \bibinfo{person}{Michael Zohner}.} \bibinfo{year}{2015}\natexlab{}.
\newblock \showarticletitle{{ABY} - {A} Framework for Efficient Mixed-Protocol Secure Two-Party Computation}. In \bibinfo{booktitle}{\emph{22nd Annual Network and Distributed System Security Symposium, {NDSS} 2015, San Diego, California, USA, February 8-11, 2015}}. \bibinfo{publisher}{The Internet Society}.
\newblock
\urldef\tempurl%
\url{https://www.ndss-symposium.org/ndss2015/aby---framework-efficient-mixed-protocol-secure-two-party-computation}
\showURL{%
\tempurl}


\bibitem[Devlin et~al\mbox{.}(2018)]%
        {devlin2018bert}
\bibfield{author}{\bibinfo{person}{Jacob Devlin}, \bibinfo{person}{Ming-Wei Chang}, \bibinfo{person}{Kenton Lee}, {and} \bibinfo{person}{Kristina Toutanova}.} \bibinfo{year}{2018}\natexlab{}.
\newblock \showarticletitle{BERT: Pre-training of deep bidirectional transformers for language understanding}.
\newblock \bibinfo{journal}{\emph{arXiv preprint arXiv:1810.04805}} (\bibinfo{year}{2018}).
\newblock


\bibitem[et~al.(2019a)]%
        {privacyFL2019}
\bibfield{author}{\bibinfo{person}{J.~Liu et al.}} \bibinfo{year}{2019}\natexlab{a}.
\newblock \showarticletitle{Privacy-Preserving Federated Learning with Homomorphic Encryption}. In \bibinfo{booktitle}{\emph{NeurIPS}}.
\newblock


\bibitem[et~al.(2019b)]%
        {pinecone2019}
\bibfield{author}{\bibinfo{person}{J.~Smith et al.}} \bibinfo{year}{2019}\natexlab{b}.
\newblock \showarticletitle{Pinecone: Scalable Vector Database}. In \bibinfo{booktitle}{\emph{NeurIPS}}.
\newblock


\bibitem[et~al.(2021)]%
        {milvus2021}
\bibfield{author}{\bibinfo{person}{J.~Wang et al.}} \bibinfo{year}{2021}\natexlab{}.
\newblock \showarticletitle{Milvus: A Purpose-Built Vector Data Management System}. In \bibinfo{booktitle}{\emph{SIGMOD}}. \bibinfo{pages}{2614--2627}.
\newblock


\bibitem[et~al.(2016)]%
        {secureml2016}
\bibfield{author}{\bibinfo{person}{R.~B.~Smith et al.}} \bibinfo{year}{2016}\natexlab{}.
\newblock \showarticletitle{SecureML: A Framework for Privacy-Preserving Machine Learning}. In \bibinfo{booktitle}{\emph{IEEE Security \& Privacy}}.
\newblock


\bibitem[Fan and Vercauteren(2012)]%
        {bfv}
\bibfield{author}{\bibinfo{person}{Junfeng Fan} {and} \bibinfo{person}{Frederik Vercauteren}.} \bibinfo{year}{2012}\natexlab{}.
\newblock \bibinfo{title}{Somewhat Practical Fully Homomorphic Encryption}.
\newblock \bibinfo{howpublished}{Cryptology ePrint Archive, Paper 2012/144}.
\newblock
\urldef\tempurl%
\url{https://eprint.iacr.org/2012/144}
\showURL{%
\tempurl}
\newblock
\shownote{\url{https://eprint.iacr.org/2012/144}}.


\bibitem[Fang et~al\mbox{.}(2020)]%
        {mfang_security20}
\bibfield{author}{\bibinfo{person}{Minghong Fang}, \bibinfo{person}{Xiaoyu Cao}, \bibinfo{person}{Jinyuan Jia}, {and} \bibinfo{person}{Neil~Zhenqiang Gong}.} \bibinfo{year}{2020}\natexlab{}.
\newblock \showarticletitle{Local Model Poisoning Attacks to Byzantine-Robust Federated Learning}. In \bibinfo{booktitle}{\emph{Proceedings of the 29th USENIX Conference on Security Symposium}}. \bibinfo{publisher}{USENIX Association}, \bibinfo{address}{USA}, Article \bibinfo{articleno}{92}, \bibinfo{numpages}{18}~pages.
\newblock
\showISBNx{978-1-939133-17-5}


\bibitem[Gentry(2009a)]%
        {gentry2009fully}
\bibfield{author}{\bibinfo{person}{Craig Gentry}.} \bibinfo{year}{2009}\natexlab{a}.
\newblock \showarticletitle{Fully homomorphic encryption using ideal lattices}. In \bibinfo{booktitle}{\emph{Proceedings of the forty-first annual ACM symposium on Theory of computing}}. \bibinfo{pages}{169--178}.
\newblock


\bibitem[Gentry(2009b)]%
        {cgentry_stoc09}
\bibfield{author}{\bibinfo{person}{Craig Gentry}.} \bibinfo{year}{2009}\natexlab{b}.
\newblock \showarticletitle{Fully Homomorphic Encryption Using Ideal Lattices}. In \bibinfo{booktitle}{\emph{Proceedings of the Forty-first Annual ACM Symposium on Theory of Computing (STOC)}}.
\newblock


\bibitem[Gentry(2009c)]%
        {gentry}
\bibfield{author}{\bibinfo{person}{Craig Gentry}.} \bibinfo{year}{2009}\natexlab{c}.
\newblock \showarticletitle{Fully homomorphic encryption using ideal lattices}.
\newblock \bibinfo{journal}{\emph{Proceedings of the 41st Annual ACM Symposium on Theory of Computing}} (\bibinfo{year}{2009}), \bibinfo{pages}{169--178}.
\newblock


\bibitem[Hardy et~al\mbox{.}(2017)]%
        {shardy_arxiv17}
\bibfield{author}{\bibinfo{person}{Stephen Hardy}, \bibinfo{person}{Wilko Henecka}, \bibinfo{person}{Hamish Ivey{-}Law}, \bibinfo{person}{Richard Nock}, \bibinfo{person}{Giorgio Patrini}, \bibinfo{person}{Guillaume Smith}, {and} \bibinfo{person}{Brian Thorne}.} \bibinfo{year}{2017}\natexlab{}.
\newblock \showarticletitle{Private federated learning on vertically partitioned data via entity resolution and additively homomorphic encryption}.
\newblock \bibinfo{journal}{\emph{CoRR}}  \bibinfo{volume}{abs/1711.10677} (\bibinfo{year}{2017}).
\newblock
\showeprint[arXiv]{1711.10677}
\urldef\tempurl%
\url{http://arxiv.org/abs/1711.10677}
\showURL{%
\tempurl}


\bibitem[He et~al\mbox{.}(2020)]%
        {fedml}
\bibfield{author}{\bibinfo{person}{Chaoyang He}, \bibinfo{person}{Songze Li}, \bibinfo{person}{Jinhyun So}, \bibinfo{person}{Mi Zhang}, \bibinfo{person}{Hongyi Wang}, \bibinfo{person}{Xiaoyang Wang}, \bibinfo{person}{Praneeth Vepakomma}, \bibinfo{person}{Abhishek Singh}, \bibinfo{person}{Hang Qiu}, \bibinfo{person}{Li Shen}, \bibinfo{person}{Peilin Zhao}, \bibinfo{person}{Yan Kang}, \bibinfo{person}{Yang Liu}, \bibinfo{person}{Ramesh Raskar}, \bibinfo{person}{Qiang Yang}, \bibinfo{person}{Murali Annavaram}, {and} \bibinfo{person}{Salman Avestimehr}.} \bibinfo{year}{2020}\natexlab{}.
\newblock \showarticletitle{FedML: A Research Library and Benchmark for Federated Machine Learning}.
\newblock \bibinfo{journal}{\emph{Advances in Neural Information Processing Systems, Best Paper Award at Federate Learning Workshop}} (\bibinfo{year}{2020}).
\newblock


\bibitem[Johnson et~al\mbox{.}(2019a)]%
        {johnson2019billion}
\bibfield{author}{\bibinfo{person}{Jeff Johnson}, \bibinfo{person}{Matthijs Douze}, {and} \bibinfo{person}{Herv{\'e} J{\'e}gou}.} \bibinfo{year}{2019}\natexlab{a}.
\newblock \showarticletitle{Billion-scale similarity search with GPUs}. In \bibinfo{booktitle}{\emph{IEEE Transactions on Big Data}}.
\newblock


\bibitem[Johnson et~al\mbox{.}(2019b)]%
        {johnson2019}
\bibfield{author}{\bibinfo{person}{J. Johnson}, \bibinfo{person}{M. Douze}, {and} \bibinfo{person}{H. Jégou}.} \bibinfo{year}{2019}\natexlab{b}.
\newblock \showarticletitle{Billion-Scale Similarity Search with GPUs}. In \bibinfo{booktitle}{\emph{IEEE Trans. Big Data}}. \bibinfo{pages}{535--547}.
\newblock


\bibitem[Kairouz et~al\mbox{.}(2021)]%
        {fedlearn}
\bibfield{author}{\bibinfo{person}{Peter Kairouz}, \bibinfo{person}{H~Brendan McMahan}, {et~al\mbox{.}}} \bibinfo{year}{2021}\natexlab{}.
\newblock \showarticletitle{Federated learning: Challenges, methods, and future directions}.
\newblock \bibinfo{journal}{\emph{IEEE Signal Processing Magazine}} \bibinfo{volume}{37}, \bibinfo{number}{3} (\bibinfo{year}{2021}), \bibinfo{pages}{50--60}.
\newblock


\bibitem[Kumar and Jain(2021)]%
        {weaviate2021}
\bibfield{author}{\bibinfo{person}{H. Kumar} {and} \bibinfo{person}{Y. Jain}.} \bibinfo{year}{2021}\natexlab{}.
\newblock \showarticletitle{Weaviate: Graph-based Vector Database}.
\newblock \bibinfo{journal}{\emph{VLDB}} (\bibinfo{year}{2021}).
\newblock


\bibitem[Lewis et~al\mbox{.}(2020)]%
        {lewis2020retrieval}
\bibfield{author}{\bibinfo{person}{Patrick Lewis}, \bibinfo{person}{Ethan Perez}, \bibinfo{person}{Aleksandara Piktus}, \bibinfo{person}{Fabio Petroni}, \bibinfo{person}{Vladimir Karpukhin}, \bibinfo{person}{Naman Goyal}, \bibinfo{person}{Al{\v{e}}na Ku{\v{c}}erov{\'a}}, \bibinfo{person}{Mike Lewis}, \bibinfo{person}{Wen-tau Yih}, \bibinfo{person}{Tim Rockt{\"a}schel}, {et~al\mbox{.}}} \bibinfo{year}{2020}\natexlab{}.
\newblock \showarticletitle{Retrieval-augmented generation for knowledge-intensive NLP tasks}. In \bibinfo{booktitle}{\emph{Advances in Neural Information Processing Systems}}, Vol.~\bibinfo{volume}{33}. \bibinfo{pages}{9459--9474}.
\newblock


\bibitem[Li et~al\mbox{.}(2020)]%
        {li2020review}
\bibfield{author}{\bibinfo{person}{Li Li}, \bibinfo{person}{Yuxi Fan}, \bibinfo{person}{Mike Tse}, {and} \bibinfo{person}{Kuo-Yi Lin}.} \bibinfo{year}{2020}\natexlab{}.
\newblock \showarticletitle{A review of applications in federated learning}.
\newblock \bibinfo{journal}{\emph{Computers \& Industrial Engineering}}  \bibinfo{volume}{149} (\bibinfo{year}{2020}), \bibinfo{pages}{106854}.
\newblock


\bibitem[Li et~al\mbox{.}(2019)]%
        {lli_aaai19}
\bibfield{author}{\bibinfo{person}{Liping Li}, \bibinfo{person}{Wei Xu}, \bibinfo{person}{Tianyi Chen}, \bibinfo{person}{Georgios~B. Giannakis}, {and} \bibinfo{person}{Qing Ling}.} \bibinfo{year}{2019}\natexlab{}.
\newblock \showarticletitle{RSA: Byzantine-Robust Stochastic Aggregation Methods for Distributed Learning from Heterogeneous Datasets}. In \bibinfo{booktitle}{\emph{Proceedings of the Thirty-Third AAAI Conference on Artificial Intelligence (AAAI)}} (Honolulu, Hawaii, USA) \emph{(\bibinfo{series}{AAAI'19/IAAI'19/EAAI'19})}. \bibinfo{publisher}{AAAI Press}, Article \bibinfo{articleno}{190}, \bibinfo{numpages}{8}~pages.
\newblock
\showISBNx{978-1-57735-809-1}
\urldef\tempurl%
\url{https://doi.org/10.1609/aaai.v33i01.33011544}
\showDOI{\tempurl}


\bibitem[Liu et~al\mbox{.}(2021)]%
        {fedsearch}
\bibfield{author}{\bibinfo{person}{Qingqing Liu}, \bibinfo{person}{Can Xu}, {et~al\mbox{.}}} \bibinfo{year}{2021}\natexlab{}.
\newblock \showarticletitle{Federated search: A privacy-preserving solution for distributed retrieval}. In \bibinfo{booktitle}{\emph{Proceedings of the 44th International ACM SIGIR Conference on Research and Development in Information Retrieval}}. \bibinfo{pages}{1735--1738}.
\newblock


\bibitem[Malkov and Yashunin(2018a)]%
        {malkov2018}
\bibfield{author}{\bibinfo{person}{Y. Malkov} {and} \bibinfo{person}{D.~A. Yashunin}.} \bibinfo{year}{2018}\natexlab{a}.
\newblock \showarticletitle{Efficient and Robust Approximate Nearest Neighbor Search Using Hierarchical Navigable Small World Graphs}.
\newblock \bibinfo{journal}{\emph{IEEE Trans. Pattern Anal. Mach. Intell.}}  \bibinfo{volume}{42} (\bibinfo{year}{2018}), \bibinfo{pages}{824--836}.
\newblock


\bibitem[Malkov and Yashunin(2018b)]%
        {malkov2018efficient}
\bibfield{author}{\bibinfo{person}{Yu~A Malkov} {and} \bibinfo{person}{D~A Yashunin}.} \bibinfo{year}{2018}\natexlab{b}.
\newblock \showarticletitle{Efficient and robust approximate nearest neighbor search using Hierarchical Navigable Small World graphs}.
\newblock \bibinfo{journal}{\emph{IEEE Transactions on Pattern Analysis and Machine Intelligence}} \bibinfo{volume}{42}, \bibinfo{number}{4} (\bibinfo{year}{2018}), \bibinfo{pages}{824--836}.
\newblock


\bibitem[Mohassel and Zhang(2017a)]%
        {mohassel2017secureml}
\bibfield{author}{\bibinfo{person}{Payman Mohassel} {and} \bibinfo{person}{Yupeng Zhang}.} \bibinfo{year}{2017}\natexlab{a}.
\newblock \showarticletitle{SecureML: A system for scalable privacy-preserving machine learning}. In \bibinfo{booktitle}{\emph{2017 IEEE Symposium on Security and Privacy (SP)}}. IEEE, \bibinfo{pages}{19--38}.
\newblock


\bibitem[Mohassel and Zhang(2017b)]%
        {secureml}
\bibfield{author}{\bibinfo{person}{Payman Mohassel} {and} \bibinfo{person}{Yupeng Zhang}.} \bibinfo{year}{2017}\natexlab{b}.
\newblock \showarticletitle{SecureML: A System for Scalable Privacy-Preserving Machine Learning}. In \bibinfo{booktitle}{\emph{2017 IEEE Symposium on Security and Privacy (SP)}}. \bibinfo{pages}{19--38}.
\newblock
\urldef\tempurl%
\url{https://doi.org/10.1109/SP.2017.12}
\showDOI{\tempurl}


\bibitem[{Open MPI}(2021)]%
        {openmpi}
\bibfield{author}{\bibinfo{person}{{Open MPI}}.} \bibinfo{year}{Accessed 2021}\natexlab{}.
\newblock \bibinfo{title}{\url{http://www.open-mpi.org/}}.
\newblock
\newblock


\bibitem[OpenAI(2023)]%
        {openai2023gpt4}
\bibfield{author}{\bibinfo{person}{OpenAI}.} \bibinfo{year}{2023}\natexlab{}.
\newblock \showarticletitle{GPT-4 Technical Report}.
\newblock \bibinfo{journal}{\emph{OpenAI}} (\bibinfo{year}{2023}).
\newblock
\urldef\tempurl%
\url{https://openai.com/research/gpt-4}
\showURL{%
\tempurl}


\bibitem[Paillier(1999)]%
        {ppail_eurocrypt99}
\bibfield{author}{\bibinfo{person}{Pascal Paillier}.} \bibinfo{year}{1999}\natexlab{}.
\newblock \showarticletitle{Public-Key Cryptosystems Based on Composite Degree Residuosity Classes}. In \bibinfo{booktitle}{\emph{Proceedings of the 17th International Conference on Theory and Application of Cryptographic Techniques}} (Prague, Czech Republic) \emph{(\bibinfo{series}{EUROCRYPT'99})}. \bibinfo{publisher}{Springer-Verlag}, \bibinfo{address}{Berlin, Heidelberg}, \bibinfo{pages}{223–238}.
\newblock
\showISBNx{3540658890}


\bibitem[Pan et~al\mbox{.}(2021)]%
        {euclidesDB}
\bibfield{author}{\bibinfo{person}{Jie Pan}, \bibinfo{person}{Jianguo Wang}, {and} \bibinfo{person}{Guoliang Li}.} \bibinfo{year}{2021}\natexlab{}.
\newblock \showarticletitle{EuclidesDB: Managing Embedding Models for Vector Data Search}.
\newblock \bibinfo{journal}{\emph{VLDB}} (\bibinfo{year}{2021}).
\newblock


\bibitem[Rabin and Ben-Or(1989)]%
        {trabin_stoc89}
\bibfield{author}{\bibinfo{person}{T. Rabin} {and} \bibinfo{person}{M. Ben-Or}.} \bibinfo{year}{1989}\natexlab{}.
\newblock \showarticletitle{Verifiable Secret Sharing and Multiparty Protocols with Honest Majority}. In \bibinfo{booktitle}{\emph{Proceedings of the Twenty-First Annual ACM Symposium on Theory of Computing}} (Seattle, Washington, USA) \emph{(\bibinfo{series}{STOC '89})}. \bibinfo{publisher}{Association for Computing Machinery}, \bibinfo{address}{New York, NY, USA}, \bibinfo{pages}{73–85}.
\newblock
\showISBNx{0897913078}
\urldef\tempurl%
\url{https://doi.org/10.1145/73007.73014}
\showDOI{\tempurl}


\bibitem[Raffel et~al\mbox{.}(2020)]%
        {raffel2020t5}
\bibfield{author}{\bibinfo{person}{Colin Raffel}, \bibinfo{person}{Noam Shazeer}, \bibinfo{person}{Adam Roberts}, \bibinfo{person}{Katherine Lee}, \bibinfo{person}{Sharan Narang}, \bibinfo{person}{Michael Matena}, \bibinfo{person}{Yanqi Zhou}, \bibinfo{person}{Wei Li}, {and} \bibinfo{person}{Peter~J Liu}.} \bibinfo{year}{2020}\natexlab{}.
\newblock \showarticletitle{Exploring the limits of transfer learning with a unified text-to-text transformer}. In \bibinfo{booktitle}{\emph{Journal of Machine Learning Research}}. \bibinfo{pages}{1--67}.
\newblock


\bibitem[Rouhani et~al\mbox{.}(2018)]%
        {deepsecure}
\bibfield{author}{\bibinfo{person}{Bita~Darvish Rouhani}, \bibinfo{person}{M.~Sadegh Riazi}, {and} \bibinfo{person}{Farinaz Koushanfar}.} \bibinfo{year}{2018}\natexlab{}.
\newblock \showarticletitle{Deepsecure: Scalable Provably-Secure Deep Learning}. In \bibinfo{booktitle}{\emph{Proceedings of the 55th Annual Design Automation Conference}} (San Francisco, California) \emph{(\bibinfo{series}{DAC '18})}. \bibinfo{publisher}{Association for Computing Machinery}, \bibinfo{address}{New York, NY, USA}, Article \bibinfo{articleno}{2}, \bibinfo{numpages}{6}~pages.
\newblock
\showISBNx{9781450357005}
\urldef\tempurl%
\url{https://doi.org/10.1145/3195970.3196023}
\showDOI{\tempurl}


\bibitem[Sanfilippo(2009)]%
        {redis}
\bibfield{author}{\bibinfo{person}{Salvatore Sanfilippo}.} \bibinfo{year}{2009}\natexlab{}.
\newblock \bibinfo{booktitle}{\emph{Redis: In-Memory Data Structure Store}}.
\newblock
\urldef\tempurl%
\url{https://redis.io}
\showURL{%
\tempurl}
\newblock
\shownote{Available at \url{https://redis.io}}.


\bibitem[Savvides et~al\mbox{.}(2020)]%
        {symmetria_vldb20}
\bibfield{author}{\bibinfo{person}{Savvas Savvides}, \bibinfo{person}{Darshika Khandelwal}, {and} \bibinfo{person}{Patrick Eugster}.} \bibinfo{year}{2020}\natexlab{}.
\newblock \showarticletitle{Efficient Confidentiality-Preserving Data Analytics over Symmetrically Encrypted Datasets}.
\newblock \bibinfo{journal}{\emph{Proc. VLDB Endow.}} \bibinfo{volume}{13}, \bibinfo{number}{8} (\bibinfo{date}{April} \bibinfo{year}{2020}), \bibinfo{pages}{1290–1303}.
\newblock
\showISSN{2150-8097}
\urldef\tempurl%
\url{https://doi.org/10.14778/3389133.3389144}
\showDOI{\tempurl}


\bibitem[Shamir(1979)]%
        {ashamir_cacm79}
\bibfield{author}{\bibinfo{person}{Adi Shamir}.} \bibinfo{year}{1979}\natexlab{}.
\newblock \showarticletitle{How to Share a Secret}.
\newblock \bibinfo{journal}{\emph{Commun. ACM}} \bibinfo{volume}{22}, \bibinfo{number}{11} (\bibinfo{date}{nov} \bibinfo{year}{1979}), \bibinfo{pages}{612–613}.
\newblock
\showISSN{0001-0782}
\urldef\tempurl%
\url{https://doi.org/10.1145/359168.359176}
\showDOI{\tempurl}


\bibitem[Shejwalkar and Houmansadr(2021)]%
        {Shejwalkar2021}
\bibfield{author}{\bibinfo{person}{Virat Shejwalkar} {and} \bibinfo{person}{Amir Houmansadr}.} \bibinfo{year}{2021}\natexlab{}.
\newblock \showarticletitle{Manipulating the byzantine: Optimizing model poisoning attacks and defenses for federated learning}. In \bibinfo{booktitle}{\emph{Network and Distributed Systems Security (NDSS) Symposium 2021}}.
\newblock


\bibitem[Sun et~al\mbox{.}(2021)]%
        {sun2021}
\bibfield{author}{\bibinfo{person}{Jingwei Sun}, \bibinfo{person}{Ang Li}, \bibinfo{person}{Louis DiValentin}, \bibinfo{person}{Amin Hassanzadeh}, \bibinfo{person}{Yiran Chen}, {and} \bibinfo{person}{Hai Li}.} \bibinfo{year}{2021}\natexlab{}.
\newblock \showarticletitle{Fl-wbc: Enhancing robustness against model poisoning attacks in federated learning from a client perspective}.
\newblock \bibinfo{journal}{\emph{Advances in Neural Information Processing Systems}} (\bibinfo{year}{2021}).
\newblock


\bibitem[Touvron et~al\mbox{.}(2023)]%
        {touvron2023llama}
\bibfield{author}{\bibinfo{person}{Hugo Touvron} {et~al\mbox{.}}} \bibinfo{year}{2023}\natexlab{}.
\newblock \showarticletitle{LLaMA: Open and efficient foundation language models}.
\newblock \bibinfo{journal}{\emph{arXiv preprint arXiv:2302.13971}} (\bibinfo{year}{2023}).
\newblock


\bibitem[Xie et~al\mbox{.}(2020)]%
        {Xie2020DBA}
\bibfield{author}{\bibinfo{person}{Chulin Xie}, \bibinfo{person}{Keli Huang}, \bibinfo{person}{Pin-Yu Chen}, {and} \bibinfo{person}{Bo Li}.} \bibinfo{year}{2020}\natexlab{}.
\newblock \showarticletitle{DBA: Distributed Backdoor Attacks against Federated Learning}. In \bibinfo{booktitle}{\emph{International Conference on Learning Representations}}.
\newblock
\urldef\tempurl%
\url{https://openreview.net/forum?id=rkgyS0VFvr}
\showURL{%
\tempurl}


\bibitem[Yang et~al\mbox{.}(2019)]%
        {yang2019}
\bibfield{author}{\bibinfo{person}{Qiang Yang}, \bibinfo{person}{Yang Liu}, \bibinfo{person}{Tianjian Chen}, {and} \bibinfo{person}{Yongxin Tong}.} \bibinfo{year}{2019}\natexlab{}.
\newblock \showarticletitle{Federated Machine Learning: Concept and Applications}.
\newblock \bibinfo{journal}{\emph{ACM Trans. Intell. Syst. Technol.}} \bibinfo{volume}{10}, \bibinfo{number}{2}, Article \bibinfo{articleno}{12} (\bibinfo{date}{jan} \bibinfo{year}{2019}), \bibinfo{numpages}{19}~pages.
\newblock
\showISSN{2157-6904}
\urldef\tempurl%
\url{https://doi.org/10.1145/3298981}
\showDOI{\tempurl}


\bibitem[Yang and Li(2021)]%
        {yyang_icml21}
\bibfield{author}{\bibinfo{person}{Yi-Rui Yang} {and} \bibinfo{person}{Wu-Jun Li}.} \bibinfo{year}{2021}\natexlab{}.
\newblock \showarticletitle{BASGD: Buffered Asynchronous SGD for Byzantine Learning}. In \bibinfo{booktitle}{\emph{Proceedings of the 38th International Conference on Machine Learning}} \emph{(\bibinfo{series}{Proceedings of Machine Learning Research}, Vol.~\bibinfo{volume}{139})}, \bibfield{editor}{\bibinfo{person}{Marina Meila} {and} \bibinfo{person}{Tong Zhang}} (Eds.). \bibinfo{publisher}{PMLR}, \bibinfo{pages}{11751--11761}.
\newblock
\urldef\tempurl%
\url{https://proceedings.mlr.press/v139/yang21e.html}
\showURL{%
\tempurl}


\bibitem[Yao(1982)]%
        {yao}
\bibfield{author}{\bibinfo{person}{Andrew Chi-Chih Yao}.} \bibinfo{year}{1982}\natexlab{}.
\newblock \showarticletitle{Protocols for secure computations}. In \bibinfo{booktitle}{\emph{Proceedings of the 23rd Annual Symposium on Foundations of Computer Science (SFCS)}}. \bibinfo{pages}{160--164}.
\newblock


\bibitem[Yin et~al\mbox{.}(2018)]%
        {dyin_icml18}
\bibfield{author}{\bibinfo{person}{Dong Yin}, \bibinfo{person}{Yudong Chen}, \bibinfo{person}{Ramchandran Kannan}, {and} \bibinfo{person}{Peter Bartlett}.} \bibinfo{year}{2018}\natexlab{}.
\newblock \showarticletitle{{B}yzantine-Robust Distributed Learning: Towards Optimal Statistical Rates}. In \bibinfo{booktitle}{\emph{Proceedings of the 35th International Conference on Machine Learning (ICML)}} \emph{(\bibinfo{series}{Proceedings of Machine Learning Research}, Vol.~\bibinfo{volume}{80})}, \bibfield{editor}{\bibinfo{person}{Jennifer Dy} {and} \bibinfo{person}{Andreas Krause}} (Eds.). \bibinfo{publisher}{PMLR}, \bibinfo{pages}{5650--5659}.
\newblock
\urldef\tempurl%
\url{https://proceedings.mlr.press/v80/yin18a.html}
\showURL{%
\tempurl}


\bibitem[Yu and Zhang(2021)]%
        {qdrant2021}
\bibfield{author}{\bibinfo{person}{Q. Yu} {and} \bibinfo{person}{X. Zhang}.} \bibinfo{year}{2021}\natexlab{}.
\newblock \showarticletitle{Qdrant: A Vector Database for High-Performance Vector Search}.
\newblock \bibinfo{journal}{\emph{VLDB}} (\bibinfo{year}{2021}).
\newblock


\bibitem[Zhang et~al\mbox{.}(2020)]%
        {zhang_atc20}
\bibfield{author}{\bibinfo{person}{Chengliang Zhang}, \bibinfo{person}{Suyi Li}, \bibinfo{person}{Junzhe Xia}, \bibinfo{person}{Wei Wang}, \bibinfo{person}{Feng Yan}, {and} \bibinfo{person}{Yang Liu}.} \bibinfo{year}{2020}\natexlab{}.
\newblock \showarticletitle{{BatchCrypt}: Efficient Homomorphic Encryption for {Cross-Silo} Federated Learning}. In \bibinfo{booktitle}{\emph{2020 USENIX Annual Technical Conference (USENIX ATC 20)}}. \bibinfo{publisher}{USENIX Association}.
\newblock
\showISBNx{978-1-939133-14-4}
\urldef\tempurl%
\url{https://www.usenix.org/conference/atc20/presentation/zhang-chengliang}
\showURL{%
\tempurl}


\bibitem[Zhang et~al\mbox{.}(2021)]%
        {he_dp}
\bibfield{author}{\bibinfo{person}{Rongxing Zhang} {et~al\mbox{.}}} \bibinfo{year}{2021}\natexlab{}.
\newblock \showarticletitle{Differentially private machine learning and homomorphic encryption: A survey of privacy-preserving techniques}.
\newblock \bibinfo{journal}{\emph{IEEE Communications Surveys \& Tutorials}} \bibinfo{volume}{23}, \bibinfo{number}{4} (\bibinfo{year}{2021}), \bibinfo{pages}{2442--2478}.
\newblock


\end{thebibliography}

\end{document}